\documentclass[fleqn,usenatbib]{mnras}
\usepackage{newtxtext,newtxmath}
\usepackage[T1]{fontenc}
\DeclareRobustCommand{\VAN}[3]{#2}
\let\VANthebibliography\thebibliography
\def\thebibliography{\DeclareRobustCommand{\VAN}[3]{##3}\VANthebibliography}

\usepackage{graphicx}	
\usepackage{amsmath}	
\usepackage{tabularx}

\newcommand{\orcid}[1]{\href{https://orcid.org/#1}{\includegraphics[scale=0.08]{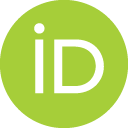}}}

\usepackage[dvipsnames]{xcolor}
\newcommand{\MSUN}{\rm{M}_{\odot}}

\title[Assembly histories from HSC Images]{ERGO-ML: The assembly histories of HSC galaxy images via invertible neural networks, contrastive learning, and cosmological simulations}

\author[L. Eisert et al.]{Lukas Eisert\orcid{0000-0003-3918-7995},$^{1,2,3}$\thanks{E-mail: eisert@slac.stanford.edu}
Connor Bottrell\orcid{0000-0003-4758-4501},$^{4}$ 
Annalisa Pillepich\orcid{0000-0003-1065-9274},$^{1}$ Dylan Nelson\orcid{0000-0001-8421-5890},$^{5}$
\newauthor
 Rhythm Shimakawa\orcid{0000-0003-4442-2750},$^{6,7}$ Marc Huertas-Company\orcid{0000-0002-1416-8483}$^{8}$, and Ralf S. Klessen\orcid{0000-0002-0560-3172}$^{5,9}$\\
$^{1}$ Max-Planck-Institut f{\"u}r Astronomie, K{\"o}nigstuhl 17, 69117 Heidelberg, Germany\\
$^{2}$ Kavli Institute for Particle Astrophysics and Cosmology, Department of Physics, Stanford University, Stanford, CA, USA \\
$^{3}$ SLAC National Accelerator Laboratory, 2575 Sandbox Hill Road, Menlo Park, CA 94025, USA \\
$^{4}$ International Centre for Radio Astronomy Research, University of Western Australia, Stirling Hwy, Crawley, WA 6009, Australia \\
$^{5}$ Universit\"{a}t Heidelberg, Zentrum f\"{u}r Astronomie, Institut f\"{u}r Theoretische Astrophysik, Albert-Ueberle-Str.\ 2, 69120 Heidelberg, Germany \\
$^{6}$ Waseda Institute for Advanced Study (WIAS), Waseda University, Nishi Waseda, Shinjuku, Tokyo 169-0051, Japan\\
$^{7}$ Center for Data Science, Waseda University, 1-6-1, Nishi-Waseda, Shinjuku, Tokyo 169-0051, Japan\\
$^{8}$ Departamento de Astrof\'{i}sica, Instituto de Astrof\'{i}sica de Canarias, Universidad de La Laguna, E-38200 La Laguna, Spain \\
$^{9}$ Universit\"{a}t Heidelberg, Interdisziplin\"{a}res Zentrum f\"{u}r Wissenschaftliches Rechnen, Im Neuenheimer Feld 225, 69120 Heidelberg, Germany \\
}

\date{Accepted XXX. Received YYY; in original form ZZZ}

\pubyear{2026}

\begin{document}
\label{firstpage}
\pagerange{\pageref{firstpage}--\pageref{lastpage}}
\maketitle

\begin{abstract}
In this paper of ERGO-ML (Extracting Reality from Galaxy Observables with Machine Learning), we develop a model that infers the merger/assembly histories of galaxies directly from optical images. We apply the self-supervised contrastive learning framework NNCLR (Nearest-Neighbor Contrastive Learning of visual Representations) on realistic HSC mock images (g,r,i - bands) produced from galaxies simulated within the TNG50 and TNG100 flagship runs of the IllustrisTNG project and with stellar masses of $10^{9-12} \MSUN$. The resulting representation is then used as conditional input for a cINN (conditional Invertible Neural Network) to gain posteriors for merger/assembly statistics, particularly the lookback time and stellar mass of the last major merger and the fraction of ex-situ stars.
Through validation against the ground truth available for simulated galaxies, we assess the performance of our model, achieving good accuracy in inferring the stellar ex-situ fraction ($\le \pm 10$ per cent for 80 per cent of the test sample) and the mass of the last major merger (within $\pm 0.5 \log \MSUN$ for stellar masses $>10^{9.5} \MSUN$ ). The information content about the lookback time is, instead, limited. We also successfully apply the TNG-trained model to simulated mocks from the EAGLE simulation, demonstrating that our model is applicable outside of the TNG domain. We hence use our simulation-based model to infer aspects of the history of observed galaxies, in particular for HSC images that are close to the domain of TNG ones. We recover the trend of increasing ex-situ stellar fraction with stellar mass and more spherical morphology, but we also identify a discrepancy between TNG and HSC: on average, observed galaxies generally exhibit lower ex-situ fractions. Despite challenges such as information loss (e.g. projection effects and surface brightness limits) and domain shifts (from simulations to observations), our results demonstrate the feasibility of extracting the merger past of galaxies from their optical images.
\end{abstract}




\section{Introduction}
Understanding the assembly history of galaxies is crucial for uncovering the processes driving cosmic evolution. How galaxies accumulate mass and experience mergers shapes their properties and appearance at any given point in time. Consequently, galaxy features can in principle provide key insights into their past. 
Traditional methods, such as the Tully-Fisher relation \citep{Tully}, have successfully linked observable scalar galaxy properties at fixed cosmic epochs. But how can one iterate on such relations when looking for --in-principle-- unobservable properties of a galaxy that are related to its past? 

Modern cosmological simulations like IllustrisTNG \citep[TNG hereafter, ][]{Springel_2017, naiman2017results, Marinacci_2018, nelson2017results, Nelson_2019, Pillepich_2018, Pillepich_2019}, EAGLE \citep{Eagle_1, Eagle_2} or SIMBA \citep{SIMBA} have produced vast numbers of galaxies whose physical states are perfectly known at each point in time. This allows astrophysicists to uncover relations, and physical connections, between observables and unobservable galaxy properties. For example, one can measure directly from the output of models the amount of stars accreted from merging, satellite and flyby galaxies, i.e. the ex-situ fractions inferred and quantified so far via semi-analytical models \citep{Huvsko_2023}, EAGLE \citep{Davison_2020}, and TNG \citep{Pillepich_2018, Boecker_2023}.
Comparisons between TNG and IFU data (e.g., MaNGA) have also been made to better understand galaxy merger histories \citep[e.g.][]{Hsu_2022, Ling_2022, Cannarozzo_2023, Zhang_2025}.

In parallel, recent advances in machine learning (ML) have opened up new ways to study galaxy evolution, particularly by allowing the learning of relations also in multi- or high-dimensional data. In our previous work \citep{Eisert_2023}, we demonstrated how simulation-based inference (SBI) using multiple observable scalar features can in principle be used to determine galaxy assembly histories. Similar approaches have been used to show how it is possible to infer the ex-situ fractions from other summary statistics of photometric images \citep{Runsheng_2025} or to directly estimate the ex-situ fraction of observed galaxies from MaNGA IFU data \citep{Angeloudi_2023, Angeloudi_2024, Angeloudi_2025}, though limited by the availability of high-quality data to nearby galaxies.

In fact, visual images represent the most accessible and most abundant form of data for studying galaxies. Survey telescopes such as SDSS, VISTA, HSC, Euclid, and Rubin capture images with very large fields of view (up to 3.5 degrees), allowing them to observe numerous galaxies simultaneously. Given the volume of data available, methods that use images are highly advantageous for large-scale, population-wide studies of galaxy evolution.

The key question of this work is the following: can the merger and assembly histories of real galaxies be inferred using visual images only? Here we focus on optical/infrared stellar light images of galaxies and aim to uncover the {\it past} history of each, and hence its integrated dynamical effects. In fact, a number of works have shown that it is possible to apply SBI to characterize the current dynamical state of galaxies \citep[][]{Bottrell_2023, Bickley_2021, Ferreiras_2020, Omori_2023, Ferreira_2024}. Here we instead aim to uncover the cumulative effects of the merger and assembly history on observed galaxies, with arguably broader prospects to understand galaxy evolution as a whole. Moreover, SBI has also already been explored to infer the star-formation history (SFH) of galaxies from their stellar spectra \citep[e.g.][and references therein]{Iglesias-Navarro_2024}: as the SFH captures the summed production of stars by a given galaxy and of all of its progenitors, it is agnostic as to how many stars were assembled and accreted via mergers and stripping. 
In contrast to SFH, stellar ex-situ fraction is an integrated summary of past mergers and assembly, and is one of the parameters we aim to infer here.

In particular, in this paper we quantitatively bridge the gap between photometric images of galaxies and their merger and assembly histories by using the output of the IllustrisTNG simulations\footnote{\url{https://www.tng-project.org/}} as ground truth and by using ML techniques to link visual features with histories. We apply the method to observed galaxies from the Hyper Suprime-Cam Subaru Strategic Program (HSC-SSP \footnote{\url{https://hsc.mtk.nao.ac.jp/ssp/}}, HSC for short hereafter). Survey and provide for the first time summary statistics of their merger and assembly histories.

To do so, we use a ResNet trained by \cite{Eisert_2024} via self-supervised contrastive learning \citep{Company_2023} to map images from various domains (from simulations: TNG, EAGLE, and SIMBA; and observations: HSC) into the same representation space. This representation space can then be used to identify galaxies that are most similar between the simulation(s) and reality as given by HSC images; in turn, such simulated galaxies from TNG can be used to train the inference model, which is then applied to observed data. Contrastive learning has recently also been successfully applied to quantitatively address the realism of simulated galaxies at the image level \citep{Eisert_2024}, to identify tidal features in HSC galaxies \citep{Desmons_2024}, as well as to study the relationship between X-ray images of the intra-cluster medium and their underlying cluster global properties \citep{Chadayammuri_2024}. However, even galaxies with similar stellar morphologies can have vastly different histories \citep[even in the case of Milky Way-like galaxies, e.g.][]{Sotillo-Ramos_2022}, which is why we use conditional Invertible Neural Networks (cINNs) \citep{FrEIA} to model the wide range of possible outcomes, i.e. to provide full posteriors.

We further validate our model by assessing whether merger and assembly histories can be accurately inferred for galaxies formed under different physical assumptions than in TNG (a domain shift). The precise implementation of e.g. feedback processes across TNG, EAGLE, and SIMBA result in galaxy populations that can differ significantly at varying levels of inspection \citep{Wright_2024, Ayromlou_2023}. This step is important because it is analogous to domain shift that might be expected between TNG physics and the real Universe. 

In Section~\ref{sec:data}, we provide an overview of the image and scalar data, both simulated and observed, used in this study. Section~\ref{sec:methods} introduces the architecture and training procedures employed thoughout. In Section~\ref{sec:domains}, we assess how well the images align in the learned representation space, and whether they are similar enough to enable inference from simulations to observations. Section \ref{sec:results} presents our inference results, starting with validation against the TNG ground truth, followed by validation against the ground truth from EAGLE, and concluding with the application to observed HSC-SSP galaxies.

\section{Galaxy Data}
\label{sec:data}
In this paper, we investigate the connection between galaxy structures and morphologies, represented by images in visual bands, and their merger and assembly history. To accomplish this, we utilize a comprehensive dataset consisting of simulated galaxy images from the TNG50, TNG100, EAGLE and SIMBA simulations. In fact, the goal of our work is the inference of the past histories of HSC-SSP galaxies, as imaged by the Hyper Suprime-Cam on Subaru. Therefore, the simulated galaxy images are tailored to mimic the characteristics of galaxy images captured by the latter. It is important to note that the data used in this study is akin to that employed in our previous works \citep{Eisert_2023, Eisert_2024}. Here, we provide a succinct overview of the data; for additional information, readers are encouraged to refer to the papers referenced above.

\subsection{Merger/assembly statistics}
\label{sec:properties}
\begin{table*}
	\centering
	\begin{tabularx}{\linewidth}{{>{\hsize=0.4\hsize}X>
	                               {\hsize=0.5\hsize}X>                            {\hsize=1.1\hsize}X}}
	\hline
	Name & Description & Note\\
	\hline
	&&\\
    Last Major Merger Mass & Total stellar mass of the last merger with a stellar mass ratio $\ge 1/4$ and a minimum total stellar mass of $8.5 \log \MSUN$. & The mass is defined as the maximum total stellar mass the secondary had during its lifetime. If there is no major merger, the parameter is set to the value of $10^4\,\MSUN$, i.e. below the resolution limit of the simulations.\\
	&&\\
	Last Major Merger Time & Lookback time of the last merger, i.e. most recent major merger. & The lookback time is relative to the redshift of the galaxy under consideration. If there is no major merger, the parameter is set to the unphysical value of $15$ Gyr.\\
	&&\\
	Stellar Ex-Situ Fraction & Ratio between ex-situ to ex-situ+in-situ stellar mass. & We track individual stellar particles through cosmic time using their unique particle ID: more specifically, we call in-situ the stars of a galaxy that have formed within its main progenitor branch and call ex-situ those that formed in galaxies on secondary branches of the galaxy merger tree, following the definitions and results of \cite{rodriguezgomez2016exsitu} and \cite{Pillepich_2018}. This mass ratio includes all stellar particles identified by \textsc{Subfind} to belong to the respective subhalo i.e. all gravitationally-bound stellar particles without using any aperture.\\
    &&\\
    \hline
	\end{tabularx}
	\caption{{\bf Unobservable galaxy properties considered in this work and inferred for observed HSC galaxies.} We list and define the unobservable properties of galaxies' past assembly and merger history that are used in this work and that are obtained by applying a ResNet and a cINN to HSC stellar-light galaxy images. Quantities are adapted from \citealt{Eisert_2023} -- note however the changed definition of a major merger, which now includes a stellar mass minimum at $8.5 \log \MSUN$ for the secondary.}
	\label{tab:properties}
\end{table*}

To summarize a galaxy's past, we select three key statistics listed in Table~\ref{tab:properties}. The definition and selection follow \cite{Eisert_2023}. However, here we choose to ignore the average merger ratio and average merger lookback time due to the relatively poor prediction performance shown in our previous study \citep{Eisert_2023} and the rather theoretical nature of average quantities derived from the simulated merger tree. Galaxy merger histories are derived from the \textsc{Sublink} tree of progenitor subhaloes/galaxies from which then meaningful scalars are derived \citep{Rodriguez_Gomez_2015}.

Cosmological simulations reveal two primary channels for star assembly: in-situ star formation and ex-situ star accretion, the latter contributing to the ex-situ stellar mass fraction distributed across galaxy bodies \citep[e.g.,][]{Oser_2010, Pillepich_2014}. Given its significance, we employ the ex-situ stellar mass fraction as a primary indicator of a galaxy's past merger history. This fraction has been previously examined in simulations \citep[e.g.,][]{rodriguezgomez2016exsitu,10.1093/mnras/staa1816} and observational discussions \citep[e.g.,][]{10.1093/mnras/stw2126}.

Major mergers (i.e. mergers with stellar mass ratios above $1/4$) can significantly influence galaxy properties. Therefore, we also consider the lookback time and stellar mass of the last major merger undergone by a galaxy \citep{rodriguezgomez2016exsitu}. We set an additional lower mass threshold for major mergers at $8.5 \log \MSUN$ in stars. Mergers falling below this threshold are in general not considered major mergers, ensuring that our sample is not contaminated by low-mass 'major mergers', particularly prevalent in TNG50 galaxies given the high resolution of the simulation.

\subsection{Observable statistics of galaxies}
\begin{table*}
	\centering
	\begin{tabularx}{\linewidth}{{> {\hsize=0.18\hsize}X>
	                          {\hsize=0.33\hsize}X>                           
                                {\hsize=0.34\hsize}X>
                                {\hsize=0.44\hsize}X}}
Input & Description  & HSC & TNG50 \& TNG100 \& EAGLE \& SIMBA\\
\hline
&&&\\
Images & 256$\times$ 256 pixels; $4 \times r_{\rm p90}$ FoV; i, r, and g bands  & see Sections~\ref{sec:images} and \ref{sec:data_preparation}& see Sections~\ref{sec:images} and \ref{sec:data_preparation}\\ 
&&&\\
Redshift & Redshift of the galaxies relative to the observer & Spectroscopic measurement \citep{Strauss_2002, GAMA_DR4} & From the discrete snapshot number/simulation-time the galaxy is taken from\\
&&&\\
$i$-band Magnitude & Apparent I-Band magnitude of stars & HSC-SSP Photometric Pipeline \citep{Bosch_2018} & Integrated stellar light incl. dust attenuation \citep{Nelson_2018} \\
&&&\\
Petrosian Radius & Radius enclosing $90$ per cent of the overall light that is originating from the respective galaxy. Determined from an isotropic Petrosian fit. & Measured via Petrofit \citep{Geda_2022} & Measured via Petrofit \\    
\hline
\end{tabularx}
	\caption{{\bf Observable galaxy properties used in this work as inputs.} We list and define the observable properties of galaxies that are used in this work as input for the ML methods and, barring the images, for matching the galaxy populations across different datasets.}
	\label{tab:observable_properties}
\end{table*}
While the main focus of this work is on inference from image data, we find that incorporating basic scalar properties as additional inputs to the models significantly improves their predictive capabilities. Specifically, we use redshift, galaxy luminosity, and radius as described in Table~\ref{tab:observable_properties}. These properties are also utilized for matching the galaxy-population sets in Section~\ref{sec:data_preparation}. Including this additional information is crucial because the galaxy images are normalized in terms of brightness and apparent size, and this normalization could result in the loss of important contextual information if the model were to rely solely on the images. 

\subsection{Galaxy-image data}
\label{sec:images}
\subsubsection{Observations: Subaru Hyper Suprime-Cam images}
We utilize galaxy images from the Public Data Release 3 (PDR3) of the Hyper Suprime-Cam Subaru Strategic Program \citep[HSC-SSP][]{HSC_DR3} as described in detail by \cite{Eisert_2024}.
The HSC-SSP, conducted over 330 nights with the Subaru telescope on Mauna Kea, Hawaii, provides a comprehensive survey with various ``layers'' of data \citep{Miyazaki_2012, Miyazaki_2018, Komiyama_2018, Furusawa_2018}. The focus here is on the Wide Layer, offering extensive coverage and depth.

To construct a galaxy sample, we use a process similar to the one outlined by \cite{Shimakawa_2022}. This involves cross-referencing sources in the HSC-SSP Wide Layer with those in the Sloan Digital Sky Survey \citep[SDSS DR16][]{SDSS_DR_16}. We apply selection criteria, including a limiting Petrosian magnitude $r_{\rm petro} < 20$ and ensuring that objects fall within the HSC-SSP footprint. Additional constraints are then used to refine the sample, removing objects affected by various factors like bright stars, bad pixels, and cosmic rays.

The galaxy sample comprises cutouts centered on each galaxy, with a consistent size and covering a range of photometric redshifts. In total, the HSC-SSP sample used in this work contains images of 768,484 galaxies in the r, g, and i bands, facilitating further analysis and investigation into galaxy properties and evolution.

\subsubsection{Simulations: TNG50 and TNG100}
We translate the stellar information content of galaxies from the TNG50 \citep{Pillepich_2019, Nelson_2019} and TNG100 \citep{Nelson_2018, Pillepich_2018, Springel_2017, naiman2017results, Marinacci_2018} simulations into synthetic observations that mimic the HSC-SSP survey.
These HSC-realistic images are described in detail by \cite{Bottrell_2023} and \cite{Eisert_2024}. 

Throughout this process, we undertake complex modeling steps: we use SKIRT for dust radiative transfer \citep{Baes_2011, Baes_2020}, which accounts for emission, absorption, and scattering of stellar light; we distinguish between light emissions from stellar populations older than 10 million years and younger populations embedded in birth clouds; and we incorporate dust attenuation by assigning dust densities to gas cells using a post-processing method proposed by \cite{Remy-Ruyer_2014} and similar to the transfer model used by \citep{Schulz_2020}.

Subsequently, we then statistically inject these images into real HSC survey images using RealSim \citep{Bottrell_2019}. This includes also physical-to-angular size scaling based on the snapshot redshift of each galaxy. It is worth noting that our study focuses on galaxies meeting specific stellar mass thresholds to ensure they contain a minimum number of stellar particles and are detectable within the sensitivity limits of the HSC-SSP.

The simulated galaxy sample is drawn from snapshots between $z=0.1$ and $z=0.4$, with stellar masses above $10^{10}~\MSUN$ (TNG100) and $10^9~\MSUN$ (TNG50). We image each galaxy along four lines of sight along each of the four vertices and the center of a tetrahedron, resulting in an extensive dataset for analysis.

\subsubsection{Simulations: EAGLE and SIMBA}
In addition to the TNG simulations, we utilize galaxy images from the EAGLE \citep{Eagle_1, Eagle_2} and SIMBA \citep{SIMBA} simulations as validation datasets for the inference models trained on TNG image data. 

For each simulation, we select galaxies with total stellar masses greater than $10^{10}~\MSUN$ at their respective $z\approx0.1$ snapshots for radiation transfer post-processing and injection into HSC images (snapshot 27 for EAGLE; 145 for SIMBA). We make mocks for $3,693$ EAGLE galaxies with $4$ camera angles each along each of the four vertices and the center of a tetrahedron and $13,293$ SIMBA galaxies in just one random camera angle, owing to its larger volume. 
The image creation pipeline for both EAGLE and SIMBA is identical to the one employed for the TNG synthetic observations, ensuring consistency in the modeling of dust radiative transfer, light emission, and integration into real HSC-SSP fields \footnote{We use versions of those simulations which have been translated into a TNG-like format by \citep{nelson2019illustristng}}. This allows for a robust comparison across different simulation sets within our ML framework.

\subsection{Data preparation}
\label{sec:data_preparation}

\begin{figure*}
	\centering
        \includegraphics[width=0.32\linewidth]{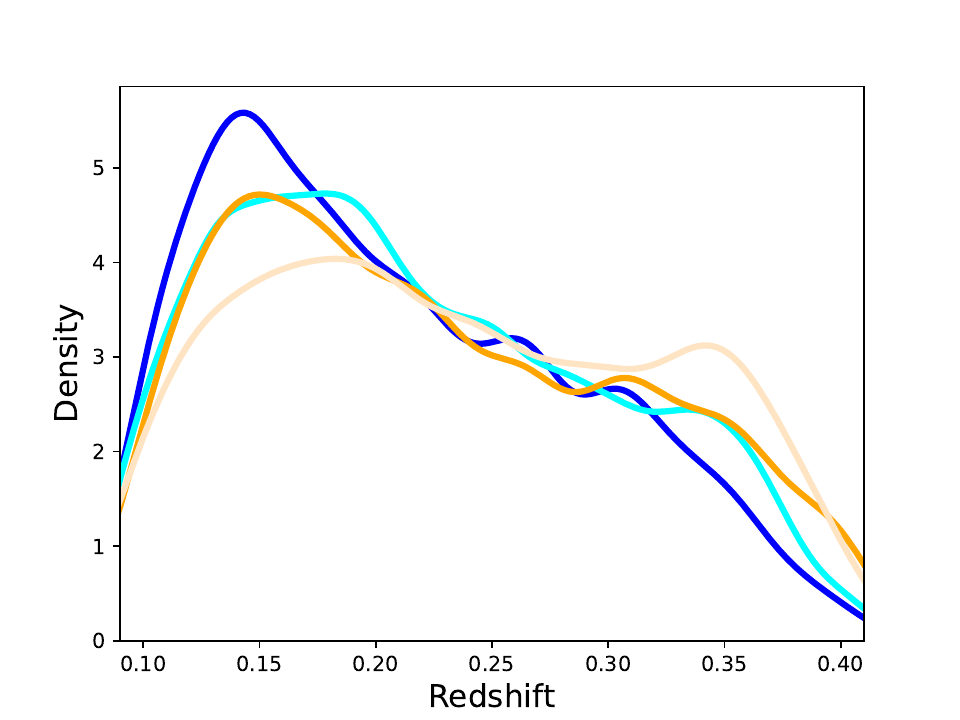}
        \includegraphics[width=0.32\linewidth]{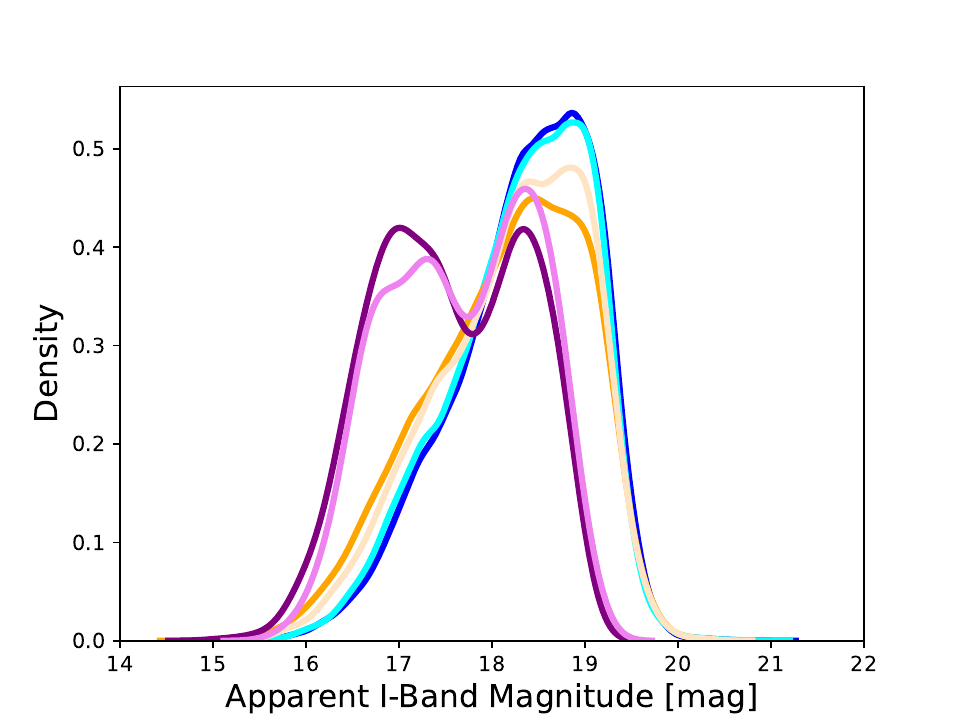}
        \includegraphics[width=0.32\linewidth]{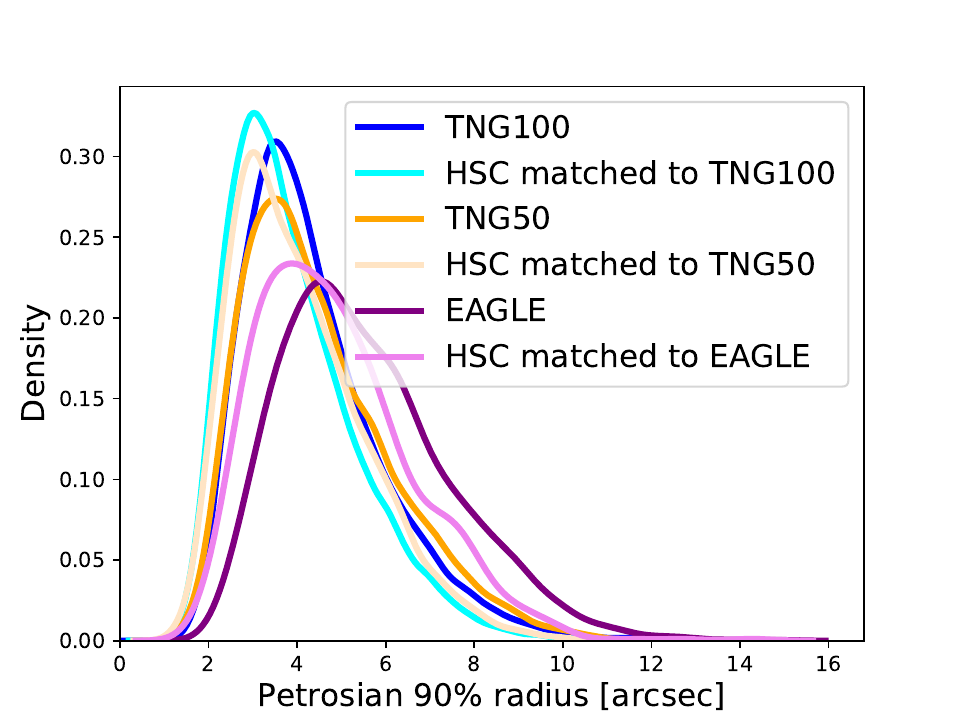}
        
	\caption{{\bf Comparison of the observed HSC dataset with the simulated galaxy images from TNG50, TNG100 and EAGLE.} The distributions of all considered datasets are shown in redshift, apparent I-band magnitude, and Petrosian radius (from left to right, see Table~\ref{tab:observable_properties} for definitions). The HSC galaxies are divided into three subsets: those matched with TNG100, with TNG50 and with EAGLE, due to differences in volume and resolution across the simulations. A larger smoothing kernel is applied to the redshift data, as the TNG simulations are available only for discrete snapshots. Galaxies with an OOD (Out of Domain) score larger than 1.2, indicating high dissimilarity, have been excluded from the analysis compared to \citet{Eisert_2024}. We find, albeit do not show, that this additional cut does not significantly impact the overall matching between the datasets. The distributions of the unmatched, parent datasets for TNG50 and TNG100 are shown in Figure 2 of \citet{Eisert_2024}. Note that the redshift distribution for EAGLE is not displayed, as the EAGLE sample used in this work consists of data from only one snapshot/redshift at $z=0.1$. Furthermore we do not display the matched distributions for SIMBA as we use those galaxies only for comparison on image realism but not for testing/validation of our models. Nevertheless, the SIMBA set undergoes the same matching procedure.}
    \label{fig:matching}
\end{figure*}

To ensure a fair comparison between real and synthetic galaxy datasets, several preparation steps are taken and described in the following Sections. 

\subsubsection{Field of View (FOV) and resizing} The FOV is based on the apparent Petrosian radius that contains $90$ per-cent of the light $r_{\rm p90}$, measured with the same algorithm for both observed and mocked images in r-band. With this, we maintain consistency across different galaxy sizes and across the domains of simulated and observed images. Images are cropped to $4 \times r_{\rm p90}$ and resized to a common grid of $256^2$ pixels.

\subsubsection{Band selection and stretching} 
RGB representation using HSC i, r, and g bands is adopted, with logarithmic image stretching to capture galaxy features across varying surface brightness levels \citep[stretching adapted from ][]{Bottrell_2019} and with all images normalized between the median of all pixels and the $99$ per-cent quantile of the central $20 \times 20$ pixels.
   
\subsubsection{Dataset splitting} The TNG50 and TNG100 simulations dataset is divided into training, validation, and test sets to prevent overfitting. Progenitor/descendant galaxies are segregated to avoid biased test results \citep{Eisert_2023}.
    
\subsubsection{Out-Of-Domain Score and image selection} 
\label{sec:ood}
We use the methodology and the representations from \cite{Eisert_2024}, based on contrastive learning, to determine whether images of simulated galaxies are {\it sufficiently realistic}: specifically, we use their Out-Of-Domain (OOD) score, which is defined and used to capture discrepancies of individual galaxies in a given domain with respect to a reference population. It is based on 8th nearest neighbour (NN) distances in representation space, which in turn is a lower-dimensional feature space in which images are encoded (see below). For a given galaxy from the ``source'' domain (e.g. a TNG galaxy), the OOD score first measures its 8th NN distance with respect to galaxies in the ``reference'' domain (e.g. HSC). This distance is then normalized by the 95th percentile of NN distances among galaxies in the reference domain. By this definition, a galaxy with an OOD score > 1.0 means that its NN distance in the reference domain is larger than 95 per-cent of the NN distances among reference domain galaxies. 

In this work, when developing the ML models to do SBI,  we neglect galaxy images with OOD > 1.2. Namely, we exclude simulated galaxies from the training sets that are significantly out-of-domain with respect to the observations, and vice versa. This approach, which is justified upon inspection of the galaxy samples (Section~\ref{sec:domains}), allows us to exclude highly unrealistic simulated galaxies that are not represented in the HSC observations, as well as HSC observations that are not well reproduced by the simulations. In this context, ``observations'' refers to the HSC image data, and ``simulations'' refers to the combined set of TNG50 and TNG100 image data.

\subsubsection{Matching of galaxy samples} Observational and simulated galaxy samples are matched based on spectroscopic redshift, apparent magnitude, and Petrosian radius to enforce a common selection function, addressing differences in distributions due to survey and simulations choices. 

The matching algorithm and results are presented in detail by \cite{Eisert_2024}. In practice, for each galaxy in a given dataset we identify suitable matches in another dataset based on similarity in redshift (with a tolerance of $\pm 0.035$), apparent i-band magnitude ($\pm 0.05$ mag) and apparent size ($\pm 0.4$ arcsec). Where no suitable match can be found, the galaxy is rejected. This matching limits our sample to the depth of the HSC sample which is $<20$ mag in the i-band.
     
\subsection{Overview of the galaxy-image samples}
Our preparation steps ensure uniformity, preventing systematic differences that could affect subsequent analysis. After matching TNG, EAGLE, and SIMBA galaxies against HSC, we have the following numbers of galaxy images:
$100,946$ (HSC); $81,901$ (TNG100); $14,947$ (TNG50); $1,566$ (EAGLE); and $2,532$ (SIMBA). Note that we remove in the matching process many galaxies from the EAGLE and SIMBA sets that are too bright and large compared to our TNG set, and this leads to the much smaller numbers compared to the raw sets.  

After additionally removing galaxies with an OOD $> 1.2$, we have a total of: $63,403$(HSC); $50,174$ (TNG100); and $10,834$ (TNG50).
These are the images used in the subsequent training and analysis of the framework. Their distributions in basic observable quantities are shown in Figure~\ref{fig:matching}. It should be kept in mind that, due to the surface brightness limit of HSC, many low-mass galaxies in TNG50 cannot be used, resulting in a simulated galaxy image set that predominantly comprises TNG100 galaxies. 

\section{From Images to Posteriors: Machine Learning Methods and Training Procedure}
\label{sec:methods}
In the past, simple relations like Tully-Fisher could be described with analytical models. However, inferring posteriors (i.e., 1D functions) for unobservable quantities from images (2D data grids) is a complex task. We use the power of neural networks to cope with this problem. 

Instead of attempting to infer posteriors of interest via convolutional neural networks starting from images directly, in this paper we introduce an intermediate step between said images (our inputs, i.e. observables) and summary statistics of the galaxies merger and assembly histories (our outputs, i.e. unobservables). Namely, we start from a ResNet trained using self-supervised contrastive learning from \cite{Eisert_2024}, which is trained to convey morphological features into a high-dimensional representation space. These representations are then here used as input for a cINN to model the posteriors of interest. We proved in previous work that this task is feasible using a limited set of integral properties \citep{Eisert_2023}, which are now replaced by the high-dimensional image representations. 

We do this contrastive learning-based intermediate step for three reasons: (1) We use the representations to test if mock images from TNG and observed HSC galaxies are similar enough to transfer the learned knowledge from TNG to HSC by calculating the distances between the image datasets in this representation space (as described in Section~\ref{sec:ood}). (2) We ensure that the training of the inference can not overfit to minor details in the simulated images (e.g. noise properties) -- the label-agnostic representations allow us to work only with relevant observable features in the images. (3) According to our tests, the inference training converges much faster and better when using pre-trained representations over an untrained CNN.

In the following, we reintroduce a few basic concepts and describe in detail the overall modeling and training procedure.

\subsection{ResNet and Contrastive Learning}
\label{sec:contrastive_resnet}
To translate the images into a low-dimensional representation that we can use as a condition for the cINNs, we employ a Wide-ResNet Architecture \citep{zagoruyko2017wide}. Within the ResNet, instead of standard convolutional layers, we use so-called "E(2)-Equivariant Steerable Convolutional" layers \citep{weiler2021general}. With this, we explicitly account for the fact that the orientation of the objects on the image plane should be irrelevant; i.e., all convolutional layers are equivariant regarding transformations of the Dihedral Group D(16). In simpler terms, this includes discrete rotations of $360^\circ / 16 = 22.5^\circ$ and reflections. Our tests showed that using dihedral groups of higher symmetries provides no additional benefit.

With the aim of speeding up training and avoiding a domain shift between observed and simulated galaxies, we train the ResNet using the self-supervised method 'Nearest Neighbor Contrastive Learning' \citep[NNCLR,][]{https://doi.org/10.48550/arxiv.2104.14548}. Here, we create two transformed images for each image and teach the ResNet to recognize transformed images of the same galaxy and discriminate between transformed images of different galaxies. The transformations we use include:
\begin{itemize}
    \item Reflections along the horizontal and vertical axes
    \item Small rotations ranging from $-10^\circ$ to $+10^\circ$
    \item Small affine shifts with a maximal displacement of $10$ per cent of the overall image side length
    \item Central zoom-ins, i.e., the center of the image stays at the center of the crop
    \item Gaussian noise
\end{itemize}
We also continue to use these transformations later during cINN training. However, for validation and testing, we use the original images.
The NNCLR method was used to pre-train the ResNet from \cite{Eisert_2024} that we use in this work but is also utilized anew in this work to ensure that the training of the cINN inference head is not leading to a domain shift when unfixing the ResNet weights. Throughout, we adopt 256 for the dimensionality of the representation space of the images.

\begin{figure*}
	\centering
        \includegraphics[trim={1cm 4cm 0cm 2cm},clip,height=18cm]{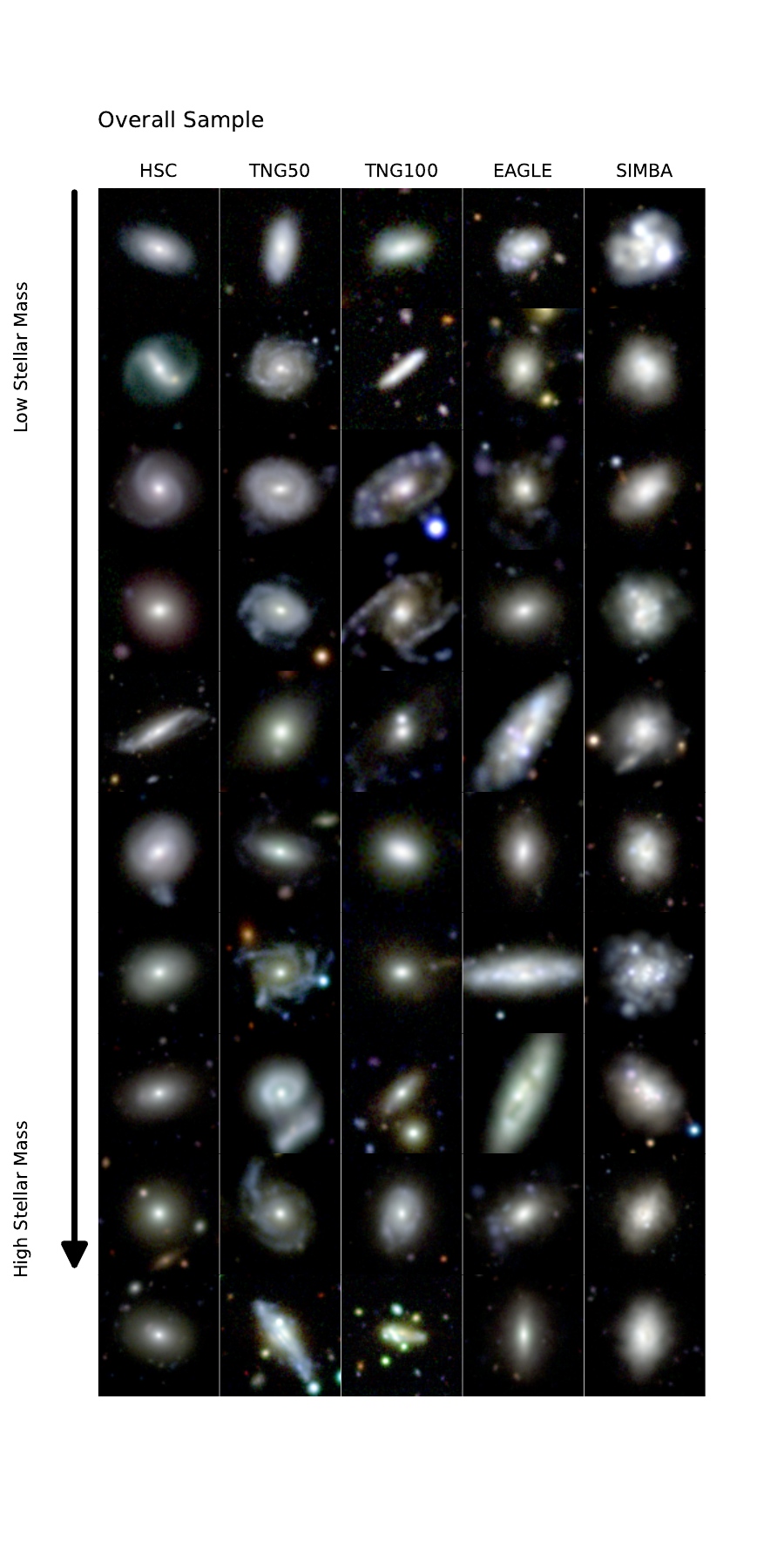}
        \includegraphics[trim={1cm 4cm 0cm 2cm},clip,height=18cm]{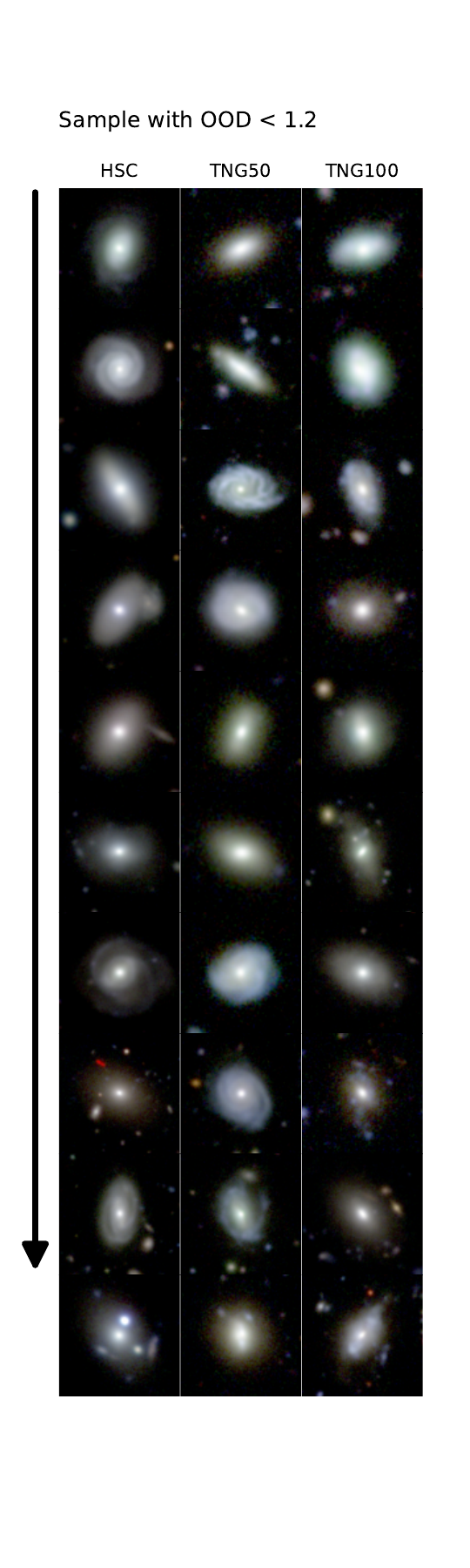}
	\caption{{\bf Overview of the galaxy images used in this work.} The left panel displays example images from the five datasets used in this work: HSC (observed) and TNG50, TNG100, EAGLE, and SIMBA (simulated). These datasets have all the same selection function in terms of redshift, Petrosian radius, and apparent i-band luminosity. The right panel shows images from the HSC, TNG50, and TNG100 samples after having applied a similarity cut (OOD score less than 1.2). These are the the types of images used for training our ML model. Each column presents 10 images, sorted into evenly spaced bins of increasing stellar mass (ranging from approximately $10^{10} \MSUN$ to $10^{12} \MSUN$). The similarity cut in the right panel visually enhances the resemblance between the images, by desire and construction. Meanwhile, diverse morphological structures are present in all considered samples.}
    \label{fig:images}
\end{figure*}

\subsection{Conditional Invertible Neural Network (cINN)}
\label{sec:cINN}
In this paper, we adopt the same architecture as by \cite{Eisert_2023}, utilizing Conditional Invertible Neural Networks (cINNs) to infer a few unobservable merger-history properties. However, unlike \cite{Eisert_2023} and as mentioned above, we employ image representations from a ResNet as conditional input. cINNs have been introduced by \cite{2019arXiv190702392A} as a generalization of Invertible Neural Networks \citep{2018arXiv180804730A}. 

These networks map, in our use case, merger-history statistics of interest $x$ to a tuple of observable image representations $y$ and latent variables $z$, offering a mechanism for inferring the posterior distribution $p(x|y)$. This is performed by enforcing a multivariate Gaussian distribution for the latent variables $z$ through negative log-likelihood loss during the training. The well-known Gaussian distribution can then be used to sample the posterior of $x$, conditioned by $y$. Implemented via the \textsc{FrEIA} framework \citep{FrEIA}, in turn based on \textsc{pytorch} \citep{NEURIPS2019_9015}, cINNs have already demonstrated success in a number of astronomical applications \citep[e.g.,][]{10.1093/mnras/staa2931, Eisert_2023}.

\subsection{Training procedure}
\label{sec:pipeline}
The overall model is represented by the cINN with the last layer of the ResNet from Section~\ref{sec:contrastive_resnet} serving as conditional input. We initialize the ResNet with the weights of the pre-trained model of \cite{Eisert_2024}. This approach allows us to begin with a network that has already demonstrated an understanding of the basic morphological features of galaxies. Furthermore, we keep the weights of the ResNet fixed in this initial step. This ensures that backpropagation focuses solely on modeling the posterior.

We utilize an Adam optimizer \citep{Kingma_2014} to back-propagate the negative log-likelihood (NLL) loss of the TNG50 and TNG100 training sets through the network. This is done in random batches of 256 images. After processing 300 randomly chosen batches (which we refer to as an epoch), we use the images in the validation set to assess the model's performance. We set an initial learning rate of 0.005, which is reduced by a factor of 0.5 if the validation loss does not decrease for 5 epochs. Once the learning rate falls below a specified level of 0.0005, we proceed to the second step of training: the ResNet weights are allowed to change now and are updated during training (they are therefore unfixed). This allows the ResNet to specialize for the inference of the quantities in question. 

Our tests indicate that significantly lower validation losses can be achieved with unfixed ResNet weights. However, there is a risk associated with unfixed ResNet weights, as the model may become overly focused on the peculiarities of the TNG mocks. To mitigate this risk, we augment the NLL loss (based on the output of the cINN) with a Contrastive Loss (based on the output of the ResNet). This contrastive loss is calculated on an additional batch of training images from the TNG50, TNG100, and HSC samples. This approach actively prevents the domain of simulated and observed galaxy images from diverging.

Therefore, the overall loss of the second step is expressed as:
\begin{align*}
L = L_{\rm NLL}(I_{\rm TNG}) + L_{\rm CL}(I_{\rm TNG}, I_{\rm HSC})
\end{align*}
where $I_{\rm TNG}$ represents the simulated TNG50 and TNG100 images, and $I_{\rm HSC}$ represents the observed HSC images. We continue the training until the validation loss $L_{\rm NLL}$ has not improved for 20 epochs.

\begin{figure*}
	\centering
	\includegraphics[trim={3cm 1cm 1.5cm 1.5cm},clip,width=0.49\linewidth]{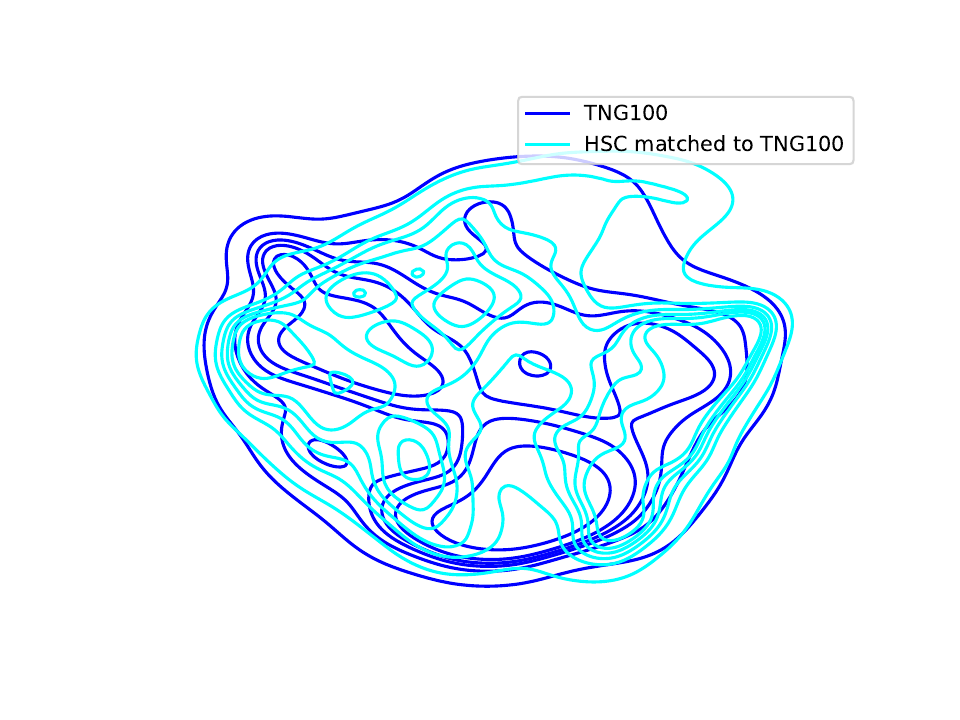}
    \includegraphics[trim={3cm 1cm 1.5cm 1.5cm},clip,width=0.49\linewidth]{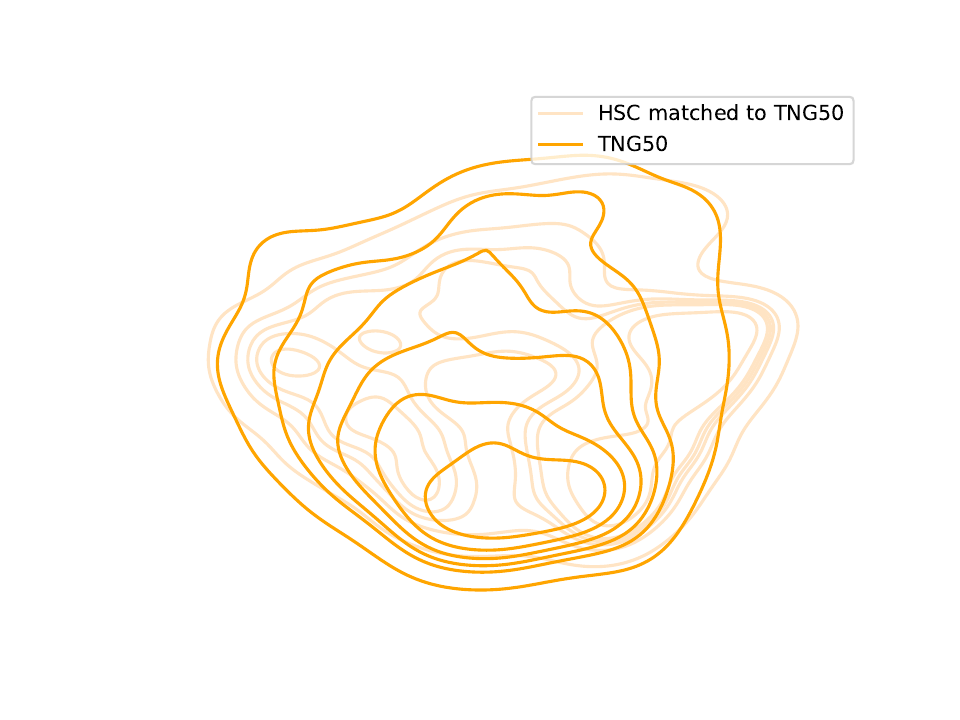}
    \includegraphics[trim={3cm 1.5cm 0.5cm 1.5cm},clip,width=0.49\linewidth]{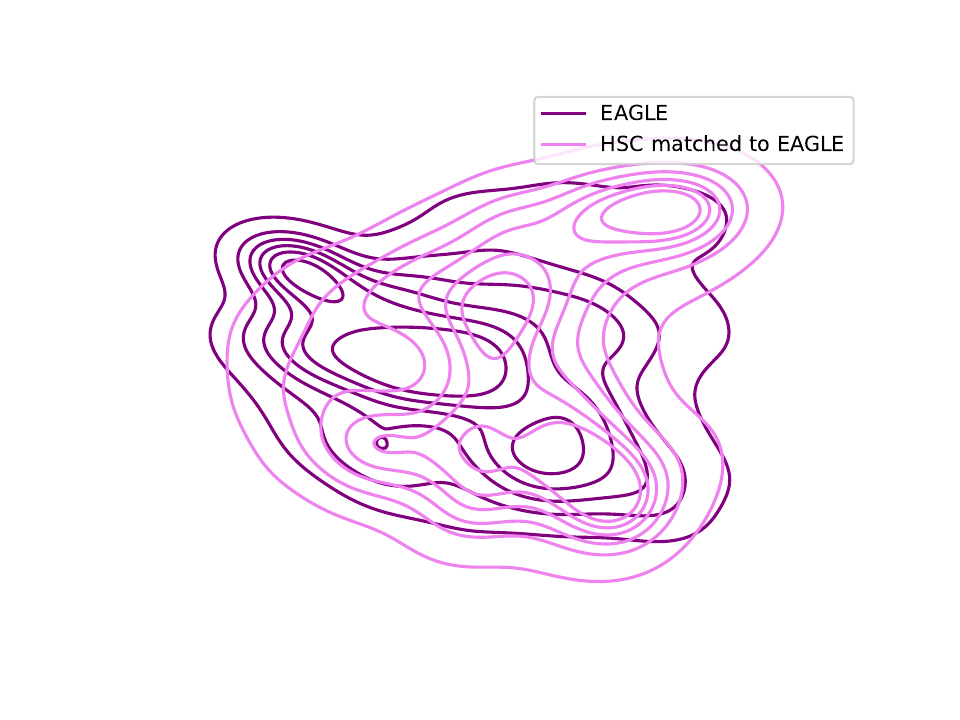}
    \includegraphics[trim={3cm 1.5cm 0.5cm 1.5cm},clip,width=0.49\linewidth]{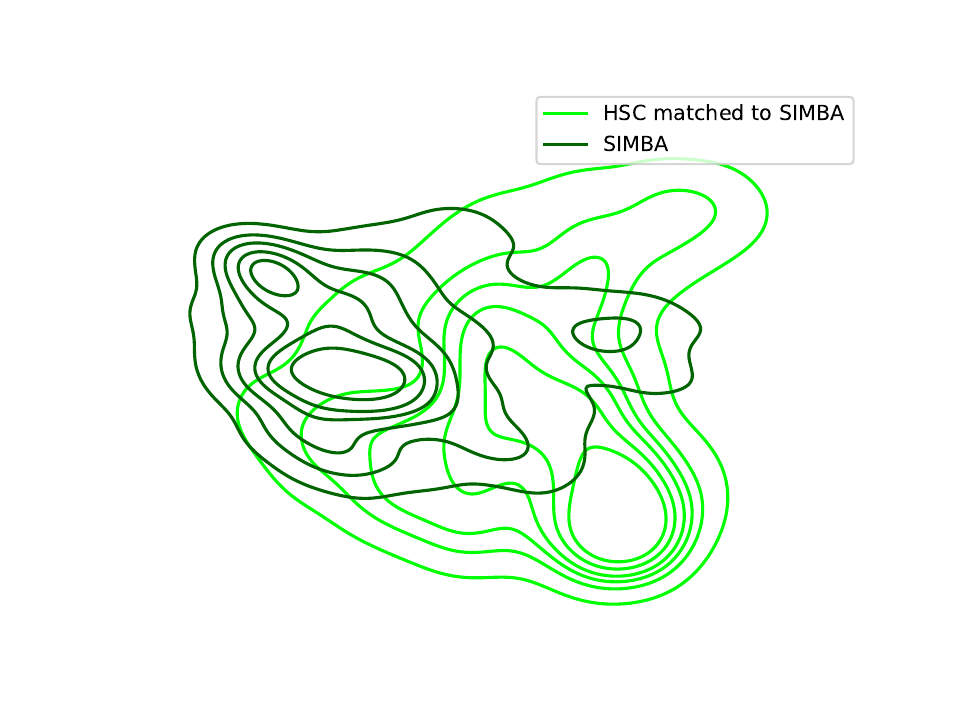}
\caption{{\bf How well do representations of the observed and simulated galaxy images align to each other?} We compare the distributions of TNG100, TNG50, EAGLE, and SIMBA images to the observed ones from HSC in the corresponding 2D-UMAP mapping of the 256-dimensional representations obtained by training a ResNet model using contrastive learning {\it simultaneously} on all  datasets. In the top-left panel, we show kernel density estimation (KDE) density plots of TNG100 images in blue and of the selection function-matched HSC set in light blue. Analogously, we compare the image density distributions of TNG50 vs. its corresponding matched HSC set (top right) and similar visualizations of the UMAPs for EAGLE (bottom left) and SIMBA (bottom right). Contours indicate isodensity lines derived from KDE in the 2D UMAP space. While there is a significant overlap among the first three sets, slight offsets and differences in point density are also evident. For the SIMBA galaxies, the distributions of images diverge most significantly. Namely, TNG100, TNG50 and EAGLE return galaxy images that are, at the population level, more consistent with observed ones from HSC than SIMBA.}
	\label{fig:umaps}
\end{figure*}

\begin{figure*}
	\centering
	\includegraphics[trim={0 0.3cm 0 1cm},clip,width=\linewidth]{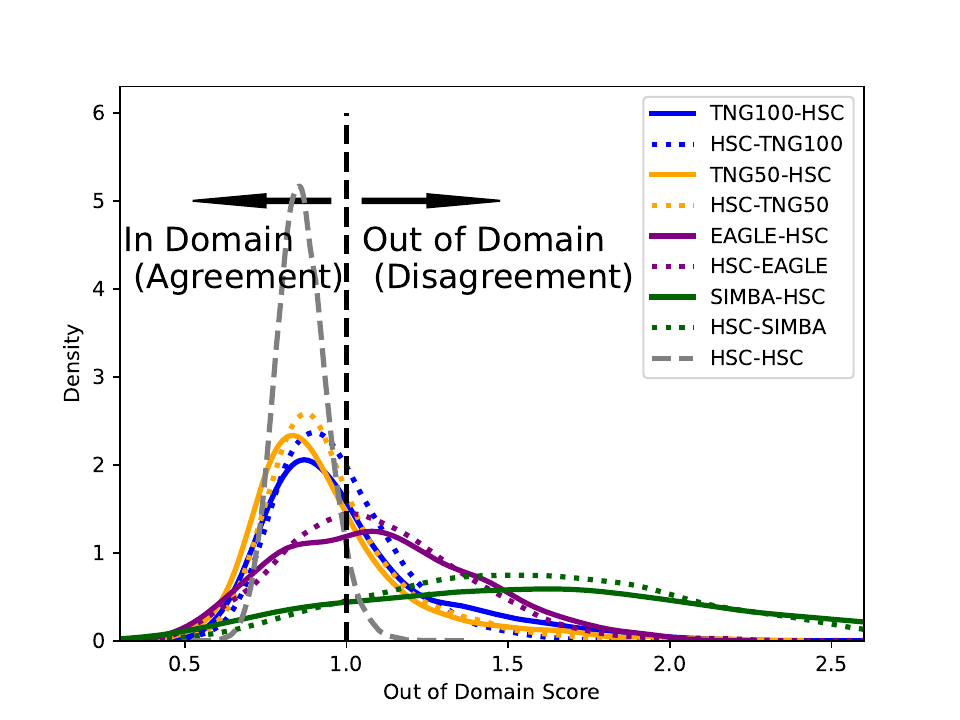}
	\caption{{\bf Out-of-Domain (OOD) Scores of the same image representations of Figure~\ref{fig:umaps}, to quantitatively assess the realism of simulated galaxies in comparison to HSC images.} We show the OOD score distributions for the observed HSC and simulated datasets (TNG50, TNG100, EAGLE, and SIMBA), following the methodology introduced by \citealt{Eisert_2024} and  using the self-supervised contrastive-learning model of Figure~\ref{fig:umaps}. The OOD score is evaluated across multiple scenarios, including comparisons between pairs of datasets and random splits within each dataset, to understand the inherent scatter: the distribution of self-distances for HSC images is provided as reference and is similar to those of the self-distances across the other simulated samples (not shown). The measurement distinguishes between the ``sides'' of the comparison, where, for example, TNG100-HSC refers to the distances of TNG100 galaxies relative to HSC, while HSC-TNG100 reflects the reverse. Images with high OOD values denote galaxies that do not resemble well observed ones. From the shape of the distributions, we can see that TNG50 and TNG100 return galaxy samples that are overall more realistic, i.e. more aligned, with HSC galaxies than EAGLE and, to a much larger degree, SIMBA. Namely, there are relatively fewer TNG50 and TNG100 galaxies than in SIMBA that appear inconsistent with HSC data.}
	\label{fig:similarity_distribution_pre_matched}
\end{figure*}

\section{The domains of the HSC, TNG, EAGLE and SIMBA Galaxy Images}
\label{sec:domains}

Before proceeding with the main goal of this paper, i.e., inferring the merger and assembly history of observed galaxies, we first assess the similarity between the training images from the simulations and the observations to which we aim to apply the model (HSC). 

\subsection{Similarity between simulated and observed galaxy samples}
The approach is the one developed by \cite{Eisert_2024}, where we have shown that, thanks to ML, it is now possible to compare observed and simulated galaxies at the map level. Crucially, and building upon such a framework, here we extend the comparison also to the images of the EAGLE and SIMBA galaxy simulations, which we use to test our model to ensure that it performs well not only on TNG galaxies but also on differently-simulated universes. 

A few example images from each of the five galaxy samples are shown on the left side of Figure~\ref{fig:images}: it gives an idea of the galaxy diversity represented in our datasets and also of the realism of the mocked images from the simulations output.

To measure the similarity among image sets, we utilize a ResNet trained using contrastive learning \citep[as in][see Section~\ref{sec:contrastive_resnet}] {Eisert_2024}, but we now do so across all datasets {\it simultaneously}: HSC, TNG50, TNG100, EAGLE, and SIMBA. This allows us to map the images into a shared 256-dimensional representation space, to then evaluate how images from the different datasets are distributed therein.

A 2D UMAP projection of these representations is displayed in Figure~\ref{fig:umaps}, whereby the contours represent varying number densities of images in the 2 dimensions of the UMAPs. We find that TNG50 and TNG100 are well aligned with the HSC sample in this space \citep[see also][]{Eisert_2024}. For EAGLE, we see overlap with the matched HSC set, though the alignment is qualitatively weaker compared to TNG. On the other hand, there is a larger discrepancy between the way SIMBA galaxies populate the representation space and their corresponding matched set of HSC galaxies. 

Following \cite{Eisert_2024} and as summarized in Section~\ref{sec:ood}, we quantify these (dis)agreements by measuring the out-of-distribution (OOD) score for each galaxy in each set, and then inspect the distributions of such pair-wise distances across all images. The results are given in Figure~\ref{fig:similarity_distribution_pre_matched}.
There we show that SIMBA galaxies, in particular, deviate considerably from the observed HSC set, in contrast to the closer alignment of TNG50 and TNG100. EAGLE galaxies fall somewhere in between. 

The qualitative and quantitative comparisons of Figures~\ref{fig:umaps} and \ref{fig:similarity_distribution_pre_matched} de facto represent a new way to assess the realism of the outcome of simulations of galaxies against observations. This methodology allows comparisons at the map or image level, without having to reduce images to summary statistics, potentially loosing information. Similar image-based comparisons between simulated and observed galaxies had been proposed in the past, with a diversity of ML-based methodology \citep[e.g.][]{Huertas_Company_2019, Zanisi_2021, Vega-Ferrero_2023}. Here for the first time we apply contrastive learning to different simulations based on different galaxy formation models, providing a way to rank them (at least in a relative sense) in terms of the realism of their galaxy morphology and appearance.

It is important to note that the image samples compared above have different sizes: namely, we use roughly 40 times more images from TNG100 than from, e.g., SIMBA. Even though we train the contrastive-learning model simultaneously across all datasets, a risk may arise that the ensuing representations are more heavily determined by the larger datasets. However, we believe that the comparison results above are robust despite the different sample sizes: in fact, we find more TNG50 galaxies with OOD scores below 1 (i.e. in agreement with HSC) than TNG100, despite the fact that the TNG50 sample is smaller, suggesting that the results of Figures~\ref{fig:umaps} and \ref{fig:similarity_distribution_pre_matched} are not purely a function of sample size.

\subsection{Selection of realistic images for the inference}
From the discussion above and from Figures~\ref{fig:umaps} and \ref{fig:similarity_distribution_pre_matched}, it is clear that all image sets, nonetheless, include galaxies with high OOD scores, i.e. that do not appear similar (and hence realistic) in comparison to observed ones. In order to implement SBI with models based on galaxy-simulation images, it would seem important to learn relationships between images and desired outputs by only considering realistic-looking galaxies.

As anticipated in Section~\ref{sec:data_preparation}, we hence impose a threshold of 1.2 on the OOD score for training our inference model. This is intended to benefit the training process by excluding TNG galaxies that are too dissimilar from HSC observations, which would eventually be excluded from the analysis. Furthermore, removing out-of-domain galaxies that are highly dissimilar may help the contrastive learning process during ResNet retraining by allowing the model to focus on more subtle differences between galaxy images.

The right panel of Figure~\ref{fig:images} shows examples of the images after excluding those with an OOD score greater than 1.2. The visual overall similarity of the remaining images is notably improved compared to the full set on the left panel. While the threshold value of 1.2 is somewhat arbitrary and chosen to balance completeness, we find that our results are robust to reasonable variations in this choice, provided that objects with very large OOD scores are excluded.

Due to the high OOD scores of SIMBA galaxies, we opt to exclude the entire SIMBA dataset from further analysis and we perform inference only on the images from TNG100, TNG50 and EAGLE.

\section{The merger and assembly histories of HSC galaxies}
\label{sec:results}
In the following, we utilize the test set defined in Section~\ref{sec:data_preparation} for multiple tasks: first, we validate the prediction capability of our model. We then investigate whether the predictions obtained by training on TNG galaxies are also transferable to survey realistic mocks of other cosmological simulations, namely mocks from EAGLE.
Finally, we apply the model to HSC galaxies identified as sufficiently similar to TNG galaxies to uncover, in part, their merger and assembly history.

\begin{figure*}
	\centering
	\includegraphics[trim={3.5cm 3cm 1cm 3cm},width=19cm]{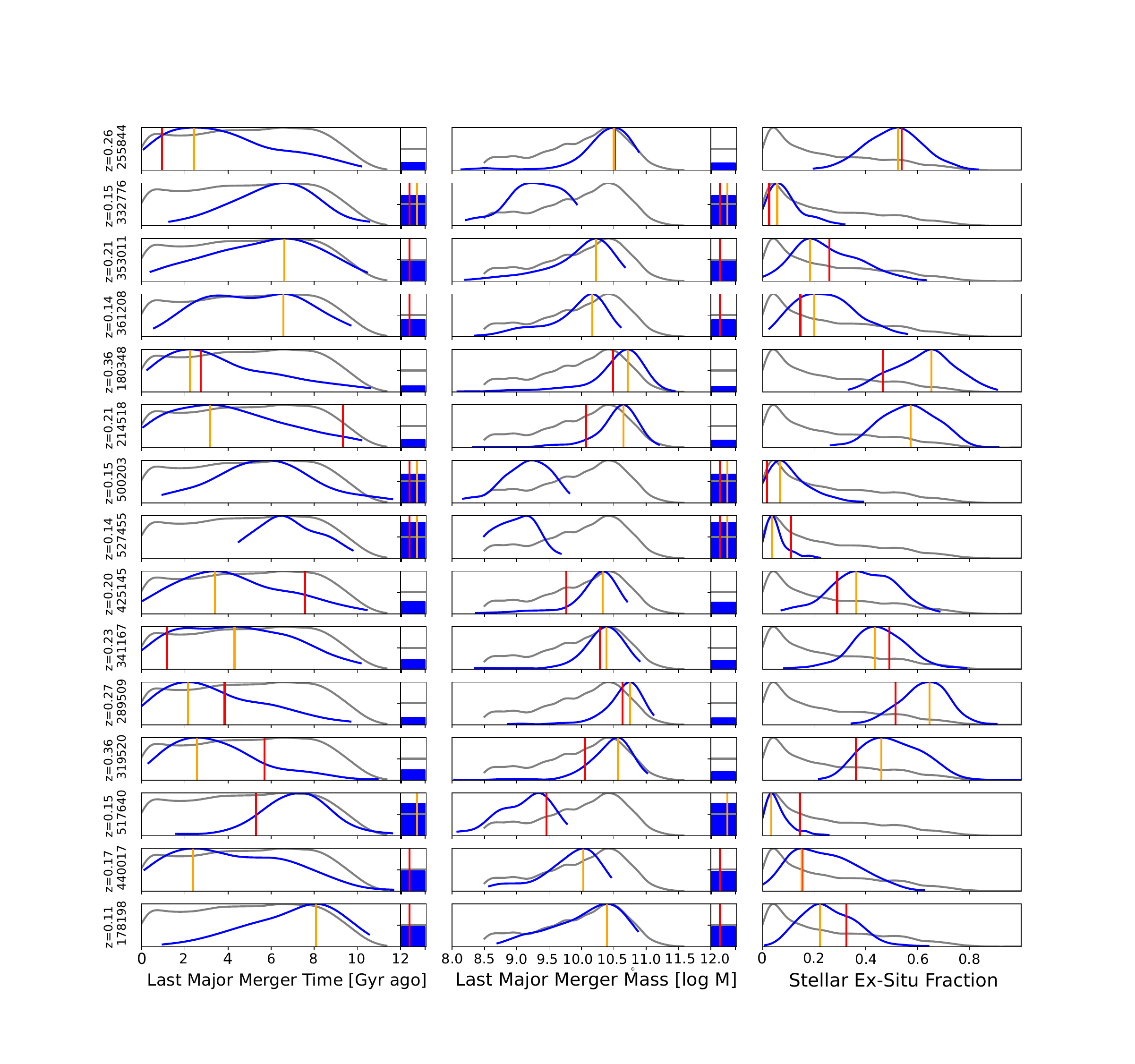}
	\caption{{\bf Posterior distributions of the three assembly history statistics inferred in this work with our SBI model and shown for 15 randomly selected galaxies from the TNG100 simulation test sets.} Each row represents a single galaxy, with its redshift and unique {\sc Subfind} ID listed on the left, while each column displays the inferred unobservable statistics of its assembly and merger history. The distributions are depicted for the entire test galaxy sample, i.e. the prior (in grey), our cINN model predictions (in blue), Maximum A-Posteriori (MAP) estimates (in yellow), and the ground truth from the TNG simulations (in red). These distributions are normalized so that the maximum value equals 1. Additionally, we include an extra bin for the last major merger quantities, indicating the fraction of posterior samples falling outside the valid regime, representing the probability that the galaxy has not experienced a major merger in its history. The MAP estimate for this bin is utilized only if the fraction of posterior samples exceeds 50 per cent, indicating a greater than 50 per cent likelihood that no major merger occurred during the lifetime of the respective galaxy.}
	\label{fig:results/TNG100/posterior_example}
\end{figure*}

\subsection{Generating posteriors and MAPs}
A cINN like the one used here (Section~\ref{sec:cINN}) is trained to provide sample points of the learned posterior distribution, collectively representing the uncertainty associated with the non-invertible nature of the problem at hand. In this work, for each galaxy in our test sample, we draw 512 points that represent this distribution.
To obtain a continuous representation of the posterior distribution from the samples, we utilize kernel density estimation (KDE). This yields a smooth, continuous function that approximates the underlying probability density function of the data. This continuous representation enables a more detailed analysis of the uncertainty associated with the inferred quantities, thereby enhancing the interpretability and utility of the model's outputs. It is important to note that for the lookback time and mass of the last major merger, we introduce an additional bin for galaxies or predictions with no major merger history (NMMs). If at least 50 per cent of posterior samples fall into this bin, the cINN MAP is set accordingly.

We display some examples of the inferred posteriors in Figure~\ref{fig:results/TNG100/posterior_example}, for a subset of test TNG100 galaxies (blue curves). From these example cases, we can see that the inference of the ex-situ stellar mass fraction appears to be quite accurate, with higher precision towards the lower ex-situ end and greater ambiguity toward larger ex-situ fractions. The lookback time of the last major mergers sometimes closely resembles the prior (gray curves), indicating that the model cannot extract any information from the image for those specific galaxies. Additionally, similar to the inference from scalar properties \citep{Eisert_2023}, we notice that the posteriors are often non-Gaussian, justifying the use of a method capable of modeling arbitrarily-shaped posteriors.

\begin{figure*}
	\centering
	\includegraphics[width=17cm]{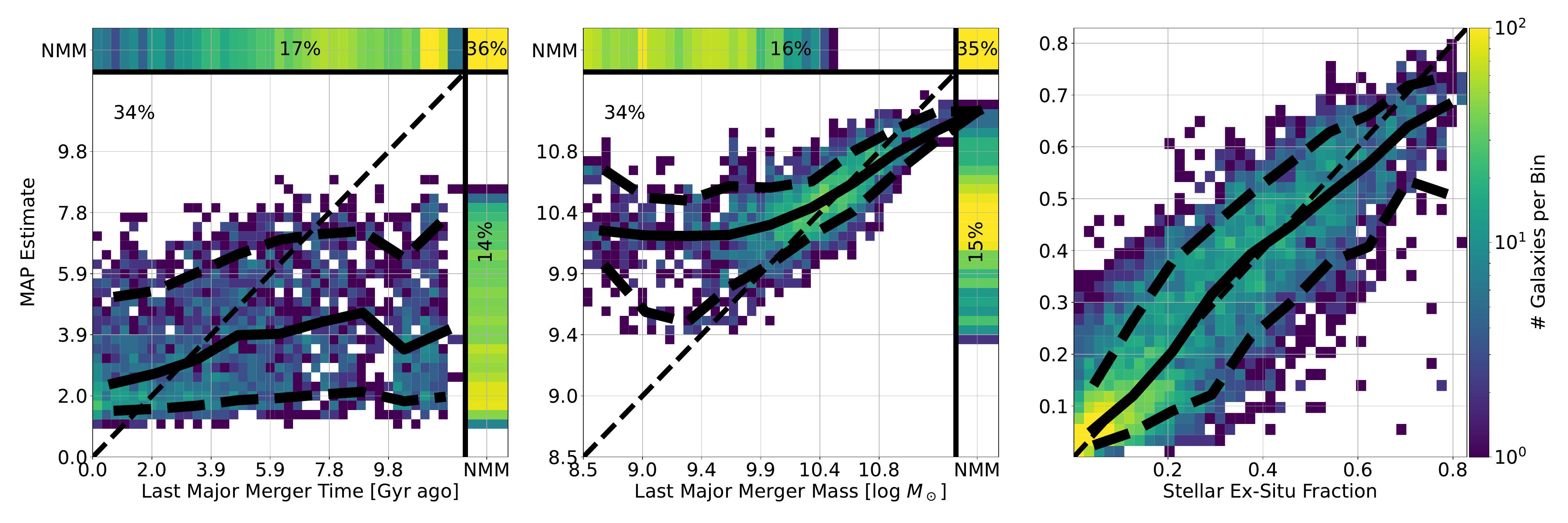}
	\includegraphics[width=17cm]{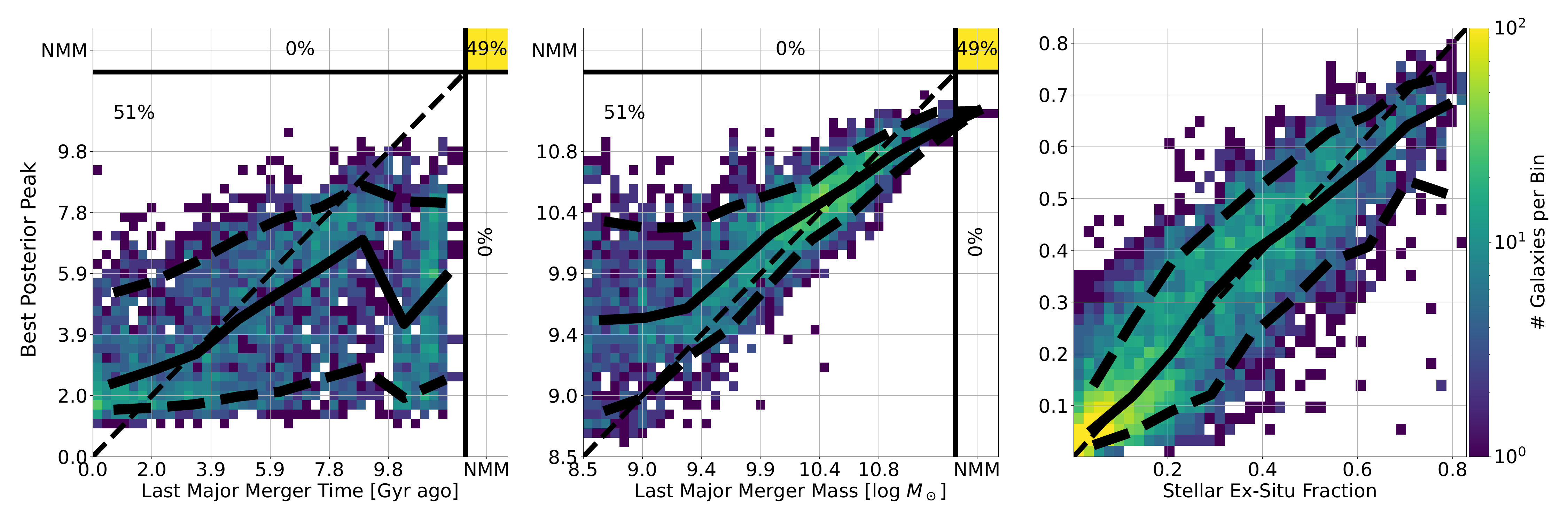}
	\caption{{\bf Validation of our cINN's predictive accuracy based on testing against ground truth data from the TNG50 and TNG100 simulations.} The columns display results for three inferred but unobservable statistics concerning the assembly and merger history of galaxies, assessed using TNG50 and TNG100 test galaxies.
    {\bf Top:} Two-dimensional histograms between cINN Maximum A-Posteriori (MAP) predictions (y-axis) and the simulation ground truth (x-axis) for each of the three output quantities. The solid black line represents the median, while the area encompassed by the black dashed lines encompasses 80 per cent of the data points. The ideal scenario, where MAP prediction and ground truth align perfectly, is depicted by a dotted black diagonal line.
    {\bf Bottom:} Similar to the top panel, but instead of MAPs, we use the inferred value of the posterior peak closest to the ground truth. This adjustment allows for the visualization of inferred values for misclassified non-major merger galaxies of the mass and lookback time of the last major merger. Additionally, for the time and mass of the last major merger (left and central panels), we show galaxies with no major merger (NMM) according to the cINN  (top), or according to the ground truth (right), and both (upper right corner). The percentage denotes the overall fraction of galaxies in each of these three categories.}
	\label{fig:results/TNG100/map}
\end{figure*}

To compare the model's predictions to the ground truth of the test TNG galaxies, we need to reduce the posterior distributions to a single quantity. While one could also use the mean or median of the posterior, we choose the MAP (Maximum A-Posteriori) estimation, which corresponds to the maximum of the posterior distribution.
To find the MAP for each unobservable quantity, we employ a KDE of the posterior based on the 512 samples. We evaluate the estimated density on a 512-point grid along each dimension. Although this approach is of discrete nature, limiting prediction accuracy, it is significantly lower than the posterior widths and therefore negligible.
To take additional information about the posterior into account, we calculate the standard deviation of the posterior point cloud along each dimension. This provides insight into the uncertainty or precision of the MAP predictions.

\subsection{Validation of the inference results}
To validate the model's predictive power on an actual population basis, we plot the MAPs defined and calculated in the previous section for the test set consisting of TNG50 and TNG100 galaxies. We then compare these predictions against the ground truth obtained from the simulations output.

The results are depicted in Figure \ref{fig:results/TNG100/map}. The upper panels show MAPs vs. ground truth where we see a very good match for the ex-situ fractions, as previously anticipated in Figure~\ref{fig:results/TNG100/posterior_example}. However, we notice that the predictions tend to favor lower ex-situ masses due to the prior, which heavily favors low ex-situ fractions. For the mass of the last major merger, we find a good match towards the high-mass end, while the predictive power decreases towards lower-mass mergers. The inferred MAPs of the times since the last major merger are very poorly consistent with the ground truth.

The fraction of correctly-identified galaxies that have had no major merger in their lifetime (or with a major merger mass below $8.5 \log \MSUN$) is 46 per cent. The fraction of falsely-identified no major merger galaxies is 20 per cent. However, we notice that the number of false classifications grows with decreasing (ground truth) major merger masses. 

In the lower panel of Figure~\ref{fig:results/TNG100/map}, we also show the results when using the peak of the posterior distribution that is closest to the ground truth as prediction. Although this is a theoretical exercise, since ground truth is unavailable for observations, it is of interest to explore the inferred lookback time and mass of the last major merger for galaxies classified as having no major merger in the upper panel. This is possible as our model generates a posterior distribution even for galaxies inferred to have no major merger. This analysis reveals a significant improvement in prediction accuracy, particularly for lower-mass major mergers and larger lookback times. 

\begin{figure*}
	\centering
	\includegraphics[width=16cm]{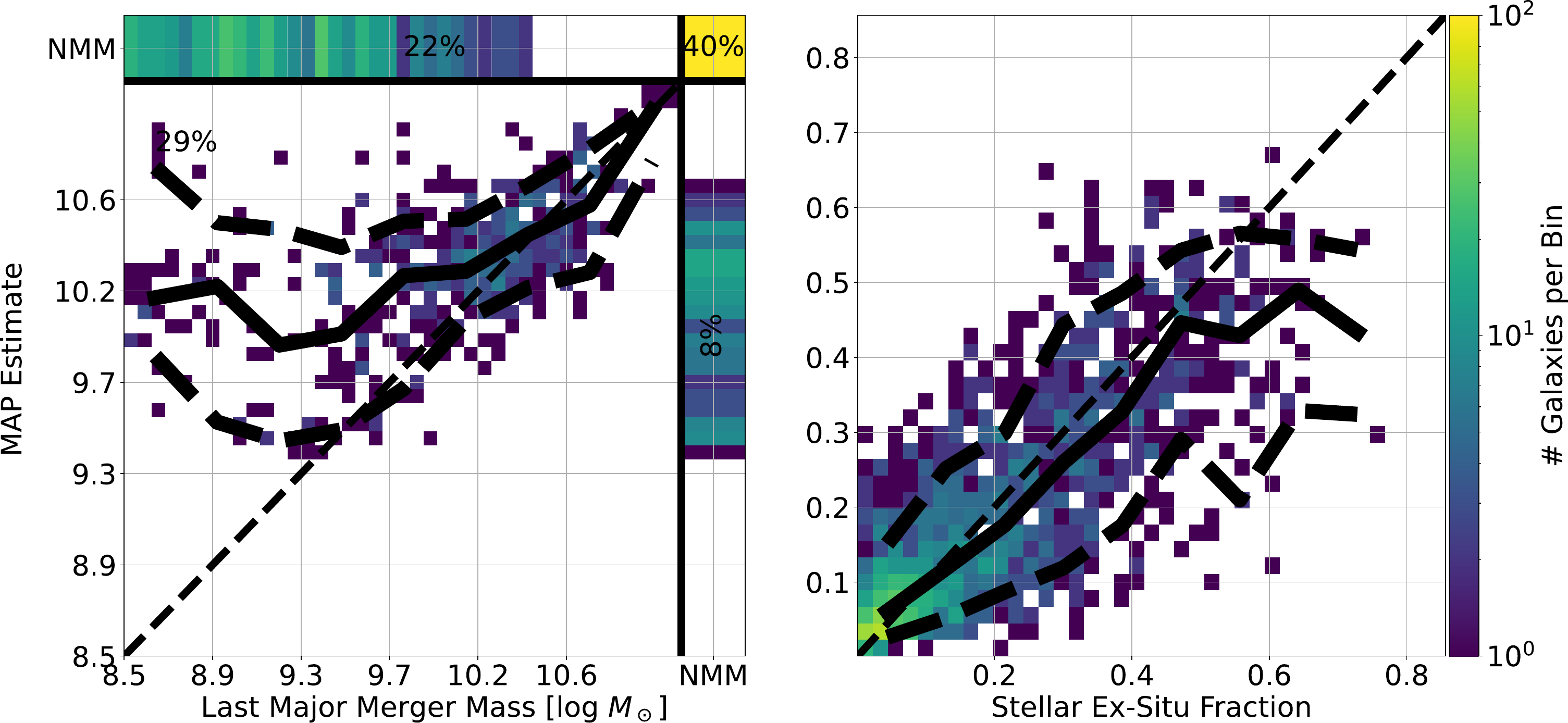}
	\caption{{\bf Validation of our cINN, trained on TNG, based on testing against ground truth data from the EAGLE simulation.} The columns display results for two inferred statistics: the stellar mass of the last major merger and the stellar ex-situ fraction. We only use EAGLE galaxies from snapshot $27$ ($z \approx 0.1$) to validate the model, and hence exclude the plot for the lookback time of the last major merger. We provide two-dimensional histograms between the cINN Maximum A-Posteriori (MAP) predictions based on TNG (y-axis) and the simulation ground truth from EAGLE (x-axis) for each of the two inferred quantities. 
    Annotations and curves are as in Figure~\ref{fig:results/TNG100/map}. We see no significant difference in prediction accuracy between applying our TNG trained model to the EAGLE Universe or to TNG one (Figure~\ref{fig:results/TNG100/map}).}
	\label{fig:map_eagle}
\end{figure*}

In the Appendix, we provide additional validation methods, including the Calibration Error (Appendix~\ref{sec:calibrationerrors}) and an analysis of the conservation of cross-correlations among merger/assembly properties (Appendix~\ref{sec:prior}). Furthermore, we compare the width of the inferred posteriors against the MAP prediction errors in Appendix~\ref{sec:uncertainties}.

\subsection{Are the results transferable from one universe to another?}

Having validated our model's predictive power on TNG50 and TNG100 galaxies, we now explore how well these results transfer to other cosmological simulations. For this task we choose to use mocks from EAGLE, as introduced in the previous Sections. This step is crucial to assess the generalizability of our model to a synthetic universe where the galaxy formation physics are comprehensive but, in detail, distinct from TNG. By extension, these simulation-to-simulation experiments should help gauging the reliability of our model when applied to the real Universe. Given that EAGLE is a distinct simulated universe with its own set of subgrid physics, testing our model on this dataset serves as a critical validation step. Importantly, since EAGLE is a simulation, we have access to the exact unobservable desired quantities, enabling a direct comparison between our model's predictions and the ground truth.  

In fact, as shown above in Figure~\ref{fig:umaps} when comparing the representations of galaxies from the EAGLE simulation with those of the observed galaxies from the HSC survey via a 2D-UMAP visualization, we find that, whereas EAGLE's representations exhibit overlap (and hence consistency) with the HSC dataset, this overlap is less pronounced compared to that between TNG and HSC.

First we identify the galaxies (observed with HSC and simulated from EAGLE) that fall into the same region in representation space as the training TNG galaxies, based on their OOD scores (Sections~\ref{sec:ood} and \ref{sec:domains}). This process is crucial to circumvent a domain shift that would render the application of the model to the domain of observed galaxies pointless. It is important to note that the representation space of the images changes during the full-inference training, as we unfreeze the ResNet weights and proceed with end-to-end training by including a contrastive-learning step, as detailed in Section~\ref{sec:pipeline}. These ``new'' OOD score distributions are shown in Figure~\ref{fig:similarity_distribution}, for comparison to those of Figure~\ref{fig:similarity_distribution_pre_matched}.

Figure~\ref{fig:map_eagle} shows that our predictions for EAGLE galaxies align well with their actual merger and assembly histories, with no significant biases or differences in comparison to the case where training and testing are done with galaxy images from the same simulation. 

This outcome is particularly exciting as it demonstrates that the model trained on TNG data is not overfitted to the specifics of the TNG simulations. Instead, it generalizes well to another cosmological simulation, EAGLE, despite the inherent differences between the two. The observational realism embedded in the images may be the reason for which the impact of these differences are mitigated, allowing the model to maintain its predictive accuracy. Interestingly, we do not find any significant trend in prediction accuracy for EAGLE galaxies with their OOD scores, i.e. their similarity to the TNG counterparts. This suggests that the model's predictive accuracy does not significantly degrade even for galaxies that lie near the edge of the domain of the images. This finding gives us confidence that our model is robust and transferable, paving the way for its application to other datasets, especially from observational surveys.

\subsection{The assembly histories of HSC galaxies from their images}
Finally, in this section, we apply the model to observed HSC galaxies. The procedure is equivalent to that in the previous section; however, we now work with real galaxies for which no ground truth is available for comparison. The validation steps taken so far give us confidence that the predictions are not subject to significant shifts. We do, however, want to emphasize that the following predictions are based on the TNG galaxy set. Moreover, given our choice of making the inference only based on images that are sufficiently realistic (Section~\ref{sec:data_preparation}), the predictions we present for observed galaxies correspond to those of similar-looking galaxies in TNG50 and TNG100. 

The tools developed in this paper, and applied to HSC galaxies, allow us to obtain the ex-situ stellar mass fraction and the mass of, and time since, the last major merger for more than 750'000 real galaxies, with related uncertainties. The corresponding catalog is made publicly available at \url{https://www.tng-project.org/data/} and it is examined and showcased in the following.

\begin{figure*}
	\centering
	\includegraphics[width=0.49\linewidth]{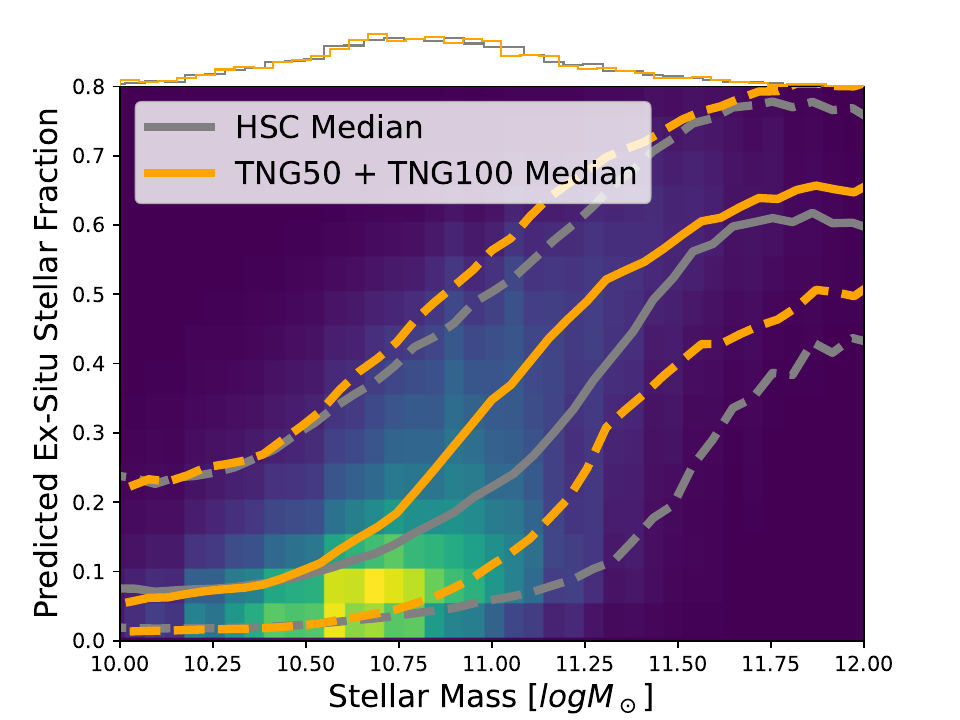}
    \includegraphics[width=0.49\linewidth]{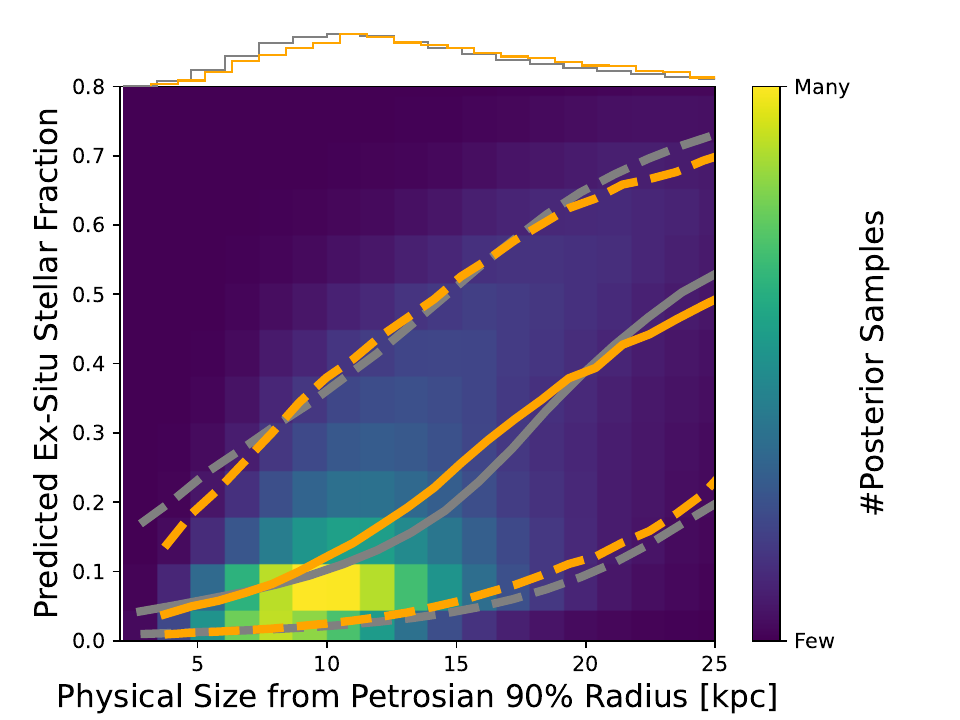}
    \includegraphics[trim={0cm 0.3cm 0cm 1cm},clip,width=\linewidth]{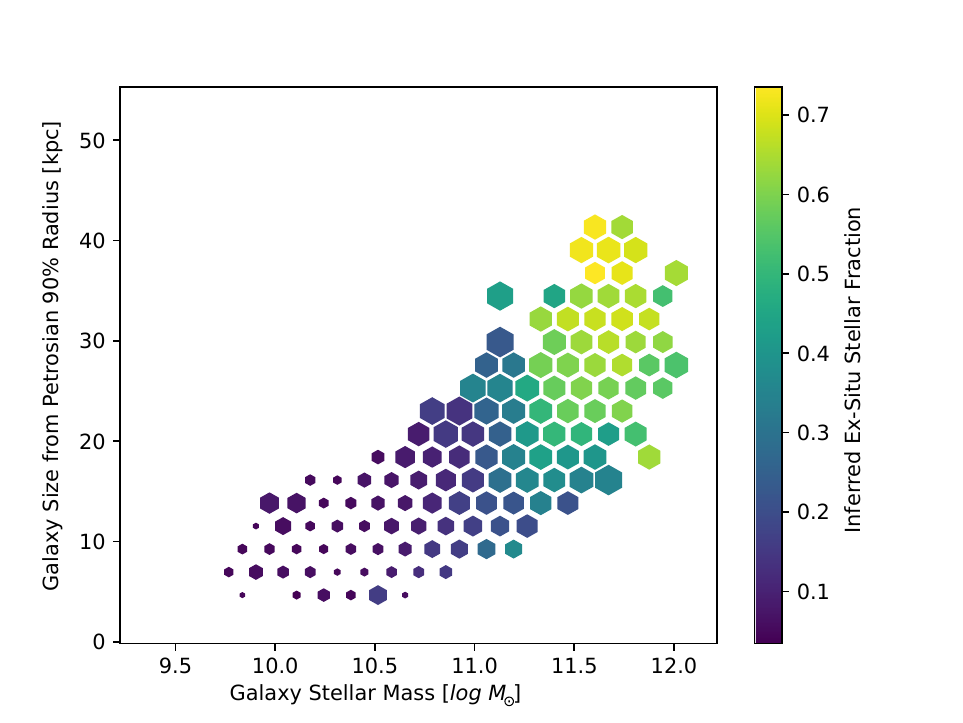}
	\caption{{\bf The ex-situ stellar mass fraction of HSC galaxies inferred by our SBI model based on the TNG50 and TNG100 simulations.} Top left: 2D histogram of the ex-situ fractions, inferred by our model, plotted against the galaxy stellar masses from the Sloan Digital Sky Survey (SDSS). For each HSC image with available SDSS stellar masses (a total of 9,043 galaxies), we plot 400 posterior samples to account for the uncertainty in predictions as modeled by our approach (color code). The median of all ex-situ fraction posterior samples in bins of stellar mass is shown for both HSC galaxies (in grey) and the combined TNG50/TNG100 dataset (in orange), in addition to dashed curves encompassing the 80 per-cent range of posterior data points in each mass bin. For the TNG galaxies, we use the total stellar mass of the corresponding galaxy. Since stellar mass measurements for HSC galaxies are available only for a subset of the overall HSC sample, we perform an additional mass matching between HSC and TNG galaxies for this plot, assuming that the two different operational definitions of a galaxy mass are consistent. Top right: same as on the left, but for the inferred ex-situ fractions plotted against galaxy stellar sizes (63,403 galaxies in total). We derive the physical radii from the Petrosian 90 per-cent light radii using the spectroscopic redshifts (Table~\ref{tab:observable_properties}) and the Petrosian radii are measured exactly in the same way in both observed and simulations maps. Main: galaxy stellar sizes vs. galaxy stellar masses of 9,043 HSC galaxies color coded by the inferred ex-situ mass fraction. Each hex-bin is colored according to the median MAP of the ex-situ posterior inferred by our model. The size of the hex-bin decode the median uncertainty (i.e. the standard deviation of the posterior) of the prediction. Small hexagons relate to a small uncertainty ($\approx 0.04$ per-cent) while large hexagons to a large uncertainty ($\approx 0.12$ per-cent). According to HSC galaxies, more massive galaxies and more extended ones (also at fixed stellar mass) are made of larger fractions of accreted stars.}
	\label{fig:inference_exsitu_mass_size}
\end{figure*}

\begin{figure*}
	\centering
	\includegraphics[width=0.49\linewidth]{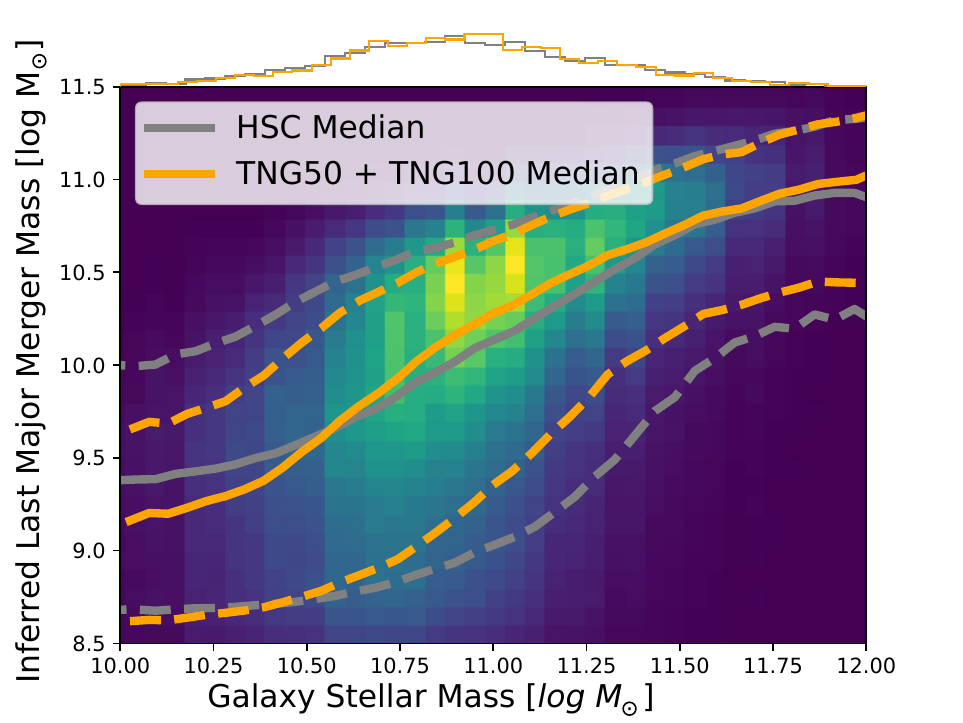}
    \includegraphics[width=0.49\linewidth]{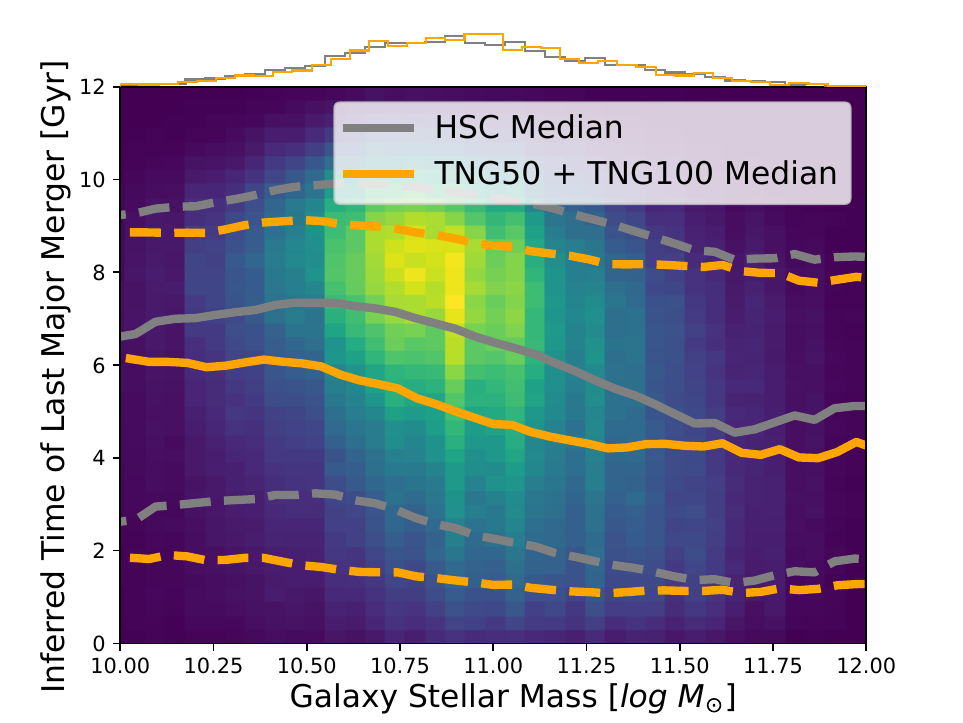}
     \includegraphics[width=0.49\linewidth]{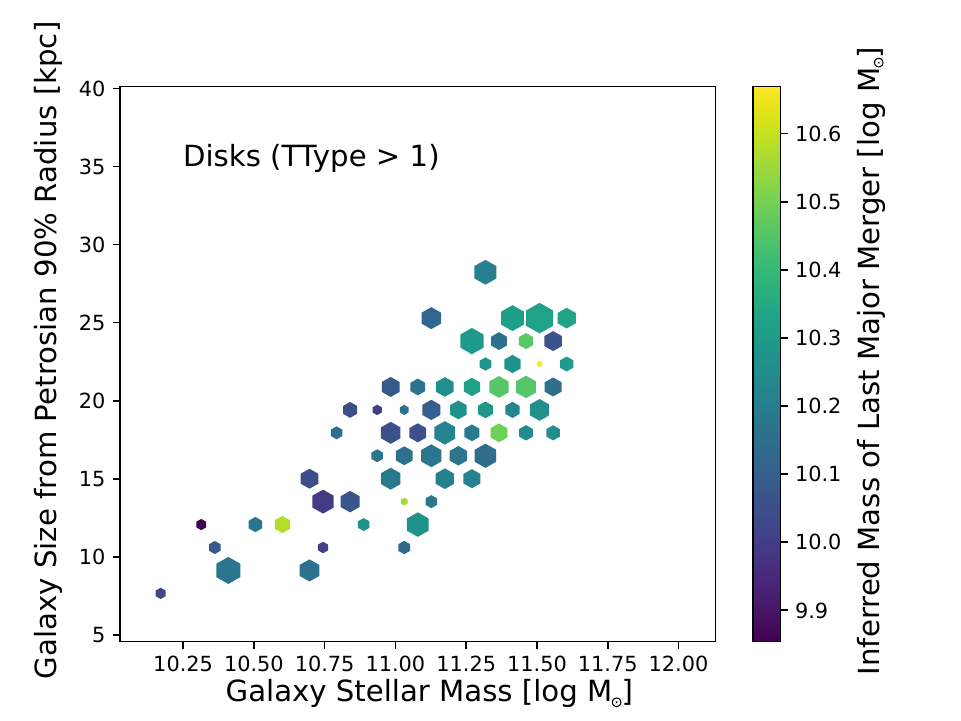}
    \includegraphics[width=0.49\linewidth]{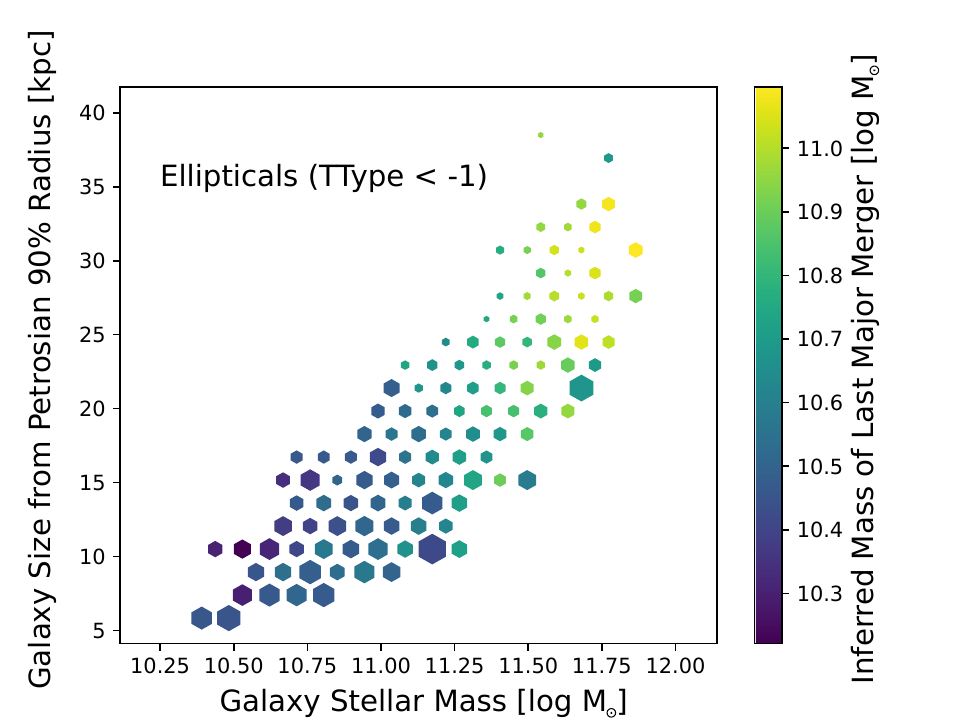}
    \includegraphics[width=0.49\linewidth]{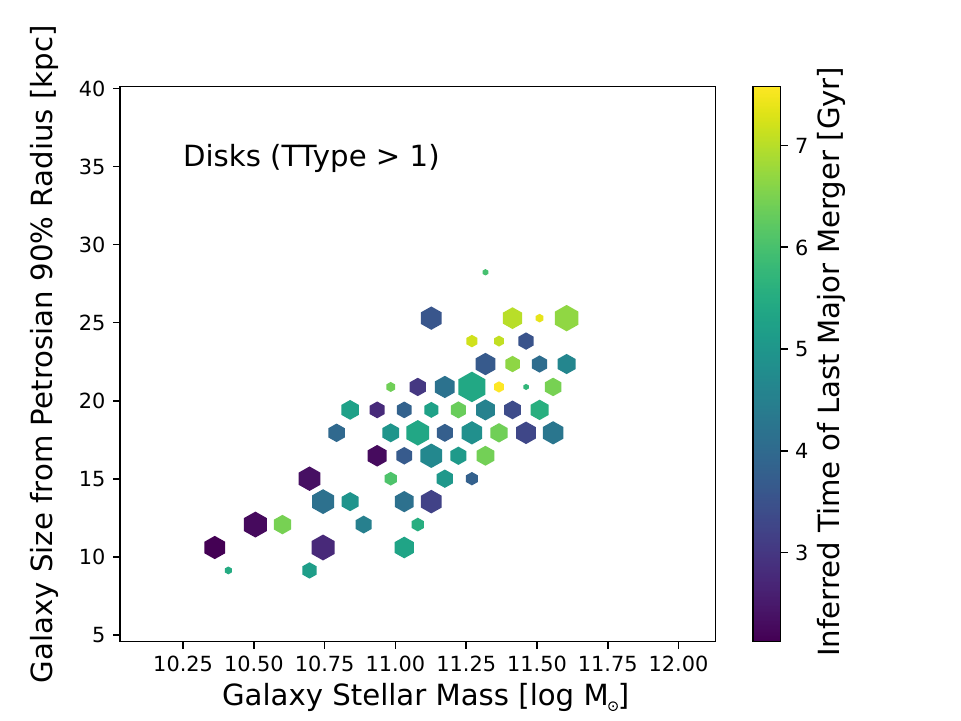}
    \includegraphics[width=0.49\linewidth]{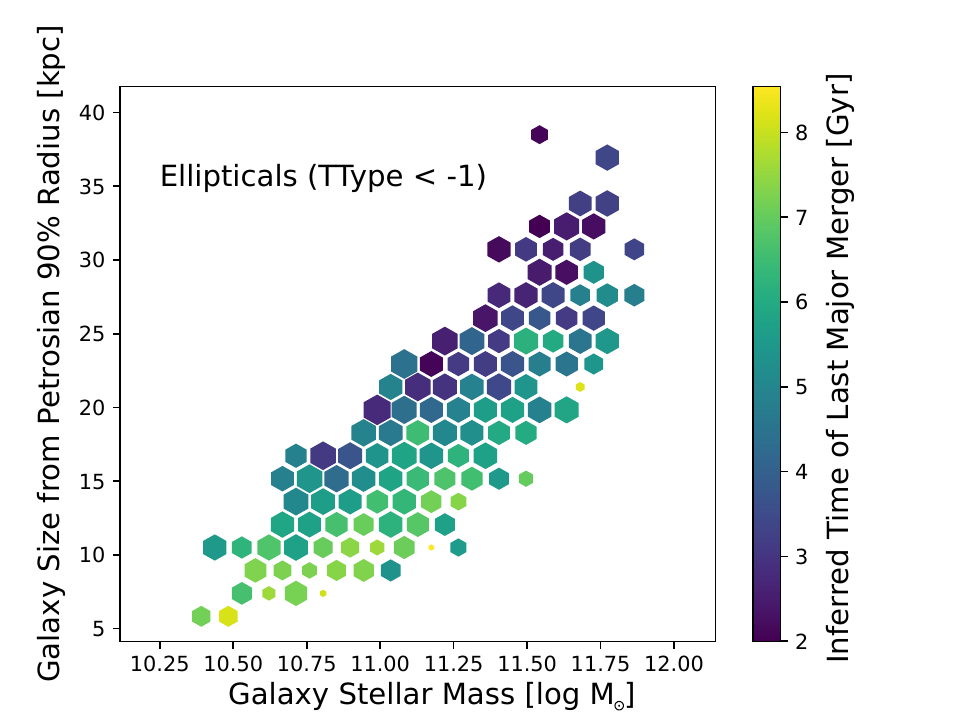}
	\caption{{\bf The mass of, and time since, the last major merger (if any) of HSC galaxies inferred by our SBI model based on the TNG50 and TNG100 simulations.} Annotations and choices are as in  Figure~\ref{fig:inference_exsitu_mass_size}.In the top two panels, for the last major merger properties as a function of galaxy stellar mass, we include all posterior samples into account: this includes posterior samples for galaxies which are inferred to have no major merger. In the lower four panels, we show the relationship among galaxy sizes, stellar masses and inferred masses/times of the last major merger for HSC galaxies split according to their stellar morphology. Whereas the time since the last major merger shows no correlation with galaxy mass, the mass of the merging galaxy clearly does so. Importantly, more massive and more extended elliptical galaxies (also at fixed stellar mass) have had more massive major mergers in the past. Despite the large uncertainties (size of the hexagons) and the poor performances of the inferences of the time since the last major merger, also the latter shows a clear relationship with galaxy sizes: at fixed stellar mass, elliptical (but not disky) HSC galaxies that are more extended have had more recent major mergers.
    }
	\label{fig:inference_mass}
\end{figure*}

\subsubsection{Larger ex-situ stellar mass fraction for more massive and more extended HSC galaxies}
\label{sec:result_1}
In Figure~\ref{fig:inference_exsitu_mass_size}, we present the inferred ex-situ stellar mass fractions for HSC galaxies in relation to their stellar mass and stellar size.

The upper left panel of Figure~\ref{fig:inference_exsitu_mass_size} shows a 2D histogram of galaxy ex-situ fractions, inferred by our model, plotted against measured stellar masses \citep[from the GALEX–SDSS–WISE Legacy Catalog 2 see][]{Salim_2016, Salim_2018}. For the 9,043 HSC galaxies with available SDSS stellar masses, we include 400 posterior samples to account for prediction uncertainty. The median ex-situ fractions for both HSC galaxies (in grey) and the combined TNG50/TNG100 dataset (in orange) are displayed, with total stellar mass used for the TNG galaxies. Ex-situ fractions rise steadily with increasing galaxy stellar mass, and a noticeable difference emerges between TNG and HSC galaxies for stellar masses between $10^{10.5}$ and $10^{11.5}\MSUN$, where HSC galaxies exhibit, on average, lower ex-situ fractions at the same stellar mass by about 10-15 percentage points. This difference will be discussed in more detail later.

The upper right panel of Figure~\ref{fig:inference_exsitu_mass_size} shows a similar trend but for ex-situ fractions against galaxy sizes. Also according to real HSC galaxies \citep[and not only according to simulated systems across galaxy-formation models, see e.g.][]{Davison_2020}, the ex-situ fraction is larger for more extended galaxies, with consistent results for simulated and observed samples.

In the main panel of Figure~\ref{fig:inference_exsitu_mass_size}, we show that, according to HSC galaxies, not only larger galaxies tend to have higher ex-situ fractions but this is the case also at fixed galaxy stellar mass \citep[as is the case for simulated galaxies, e.g.][]{Davison_2020, Zhu2021}. This figure confirms the known trend between the stellar mass and the size of a galaxy, which has been quantified for HSC galaxies by \cite{Kawinwanichakij_2021} and re-obtained here in terms of Petrosian radii. More importantly for the topics at hand, whereas the ex-situ fractions would seem primarily correlated with stellar mass, a secondary trend with galaxy size is also suggested for galaxies with stellar masses $\gtrsim10^{11} \MSUN$. In this mass range, larger galaxies have higher ex-situ fractions at fixed mass. Overall, the ex-situ fraction rises steadily with both increasing stellar mass and galaxy size in this relation.

\subsubsection{More massive last major mergers for more massive, more extended and elliptical HSC galaxies}
\label{sec:result_2}
In Figure~\ref{fig:inference_mass}, top panels, we show the predictions for the mass and the lookbacktime of the last major merger against galaxy stellar mass. For HSC, we infer  slightly earlier major mergers compared to TNG (by up to 1-2 billion years) particularly in the mass range $10^{10.5 - 11.2} \MSUN $. More massive galaxies have undergone more massive major mergers, whereas no trend emerges between galaxy stellar mass and time of the merger. Even though the positive correlation between galaxy mass and mass of the last major merger is expected within the standard cosmological scenario of cosmological structure formation, our analyses reveals that this holds true also for real observed galaxies based on their optical maps.

In the same way as more extended galaxies contain larger mass fractions of ex-situ stars, also at fixed stellar mass (Figure~\ref{fig:inference_exsitu_mass_size}), relations emerge for HSC galaxies among sizes, masses and properties of the last major merger. This is shown in the bottom four panels of Figure~\ref{fig:inference_mass}, for galaxies of different morphological types using the ML-based TType catalog for SDSS galaxies from \citep[][]{Sanchez_2018}. As our set of HSC galaxies is crossmatched against SDSS we can utilize those classifications in this study: TTypes of $-2$ denote elliptical galaxies while positive TTypes denote disks. In fact, it is known that the stellar size-mass relation is different for galaxies of different morphologies types \citep[see e.g.][for SDSS galaxies]{Shen_2003} and this is successfully returned also in the TNG simulations \citep{Genel_2018}. 

So in the bottom four panels of Figure~\ref{fig:inference_mass} we plot galaxies sizes against the stellar masses color coded by hex-bin-wise medians of the mass and time of the last major merger. 
Furthermore we split the population into disky and elliptical galaxies according to their TTypes \citep[$>1$ and $<-1$, respectively,][]{Sanchez_2018}. Also for our HSC sample we find that the size-mass relation is steeper for the elliptical population than for the disks. Strong trends with the mass and time of the last major merger (and with the ex-situ fraction, not shown) are in place -- this is the case for the elliptical population only: namely, also at fixed stellar mass, more extended {\it elliptical} galaxies have had more massive and more recent major mergers, in addition to larger ex-situ fractions. On the other hand, HSC disky galaxies do not exhibit such secondary trends.

The mass of the last major merger of elliptical galaxies is larger for more massive and more extended galaxies, and this same trend at fixed stellar mass is somewhat weak. 
Moreover, the inference uncertainty decreases significantly towards more massive and more extended ellipticals. On the other hand, and despite the large prediction errors, the time of the last major merger is encoded in the scatter of the size mass relation: at a given mass, a larger elliptical galaxy has had on average a more recent major merger.

In the disky regim, we find large ex-situ fractions $>0.2$ only for the most massive and largest galaxies. The trends of mass and lookbacktime of last major mergers are more spurious, as many of the HSC disk galaxies are inferred to have no major merger in their history.

\subsubsection{Different merger/assembly histories for HSC galaxies with different morphologies}
\label{sec:result_3}
\begin{figure}
    \centering
    \includegraphics[width=1\linewidth]{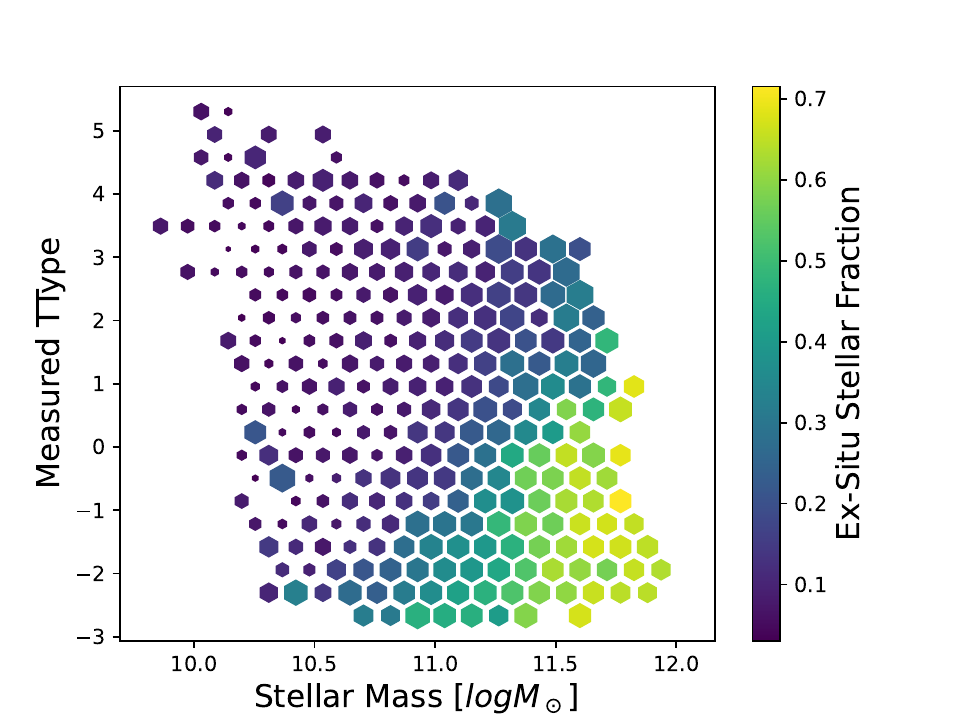}
    \includegraphics[width=1\linewidth]{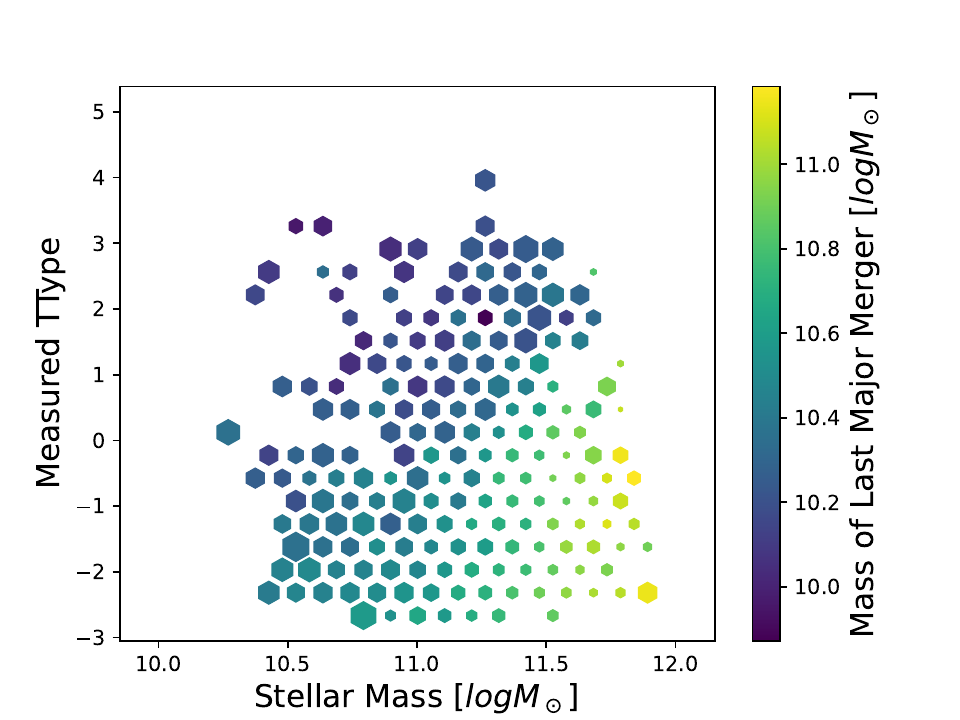}
    \includegraphics[width=1\linewidth]{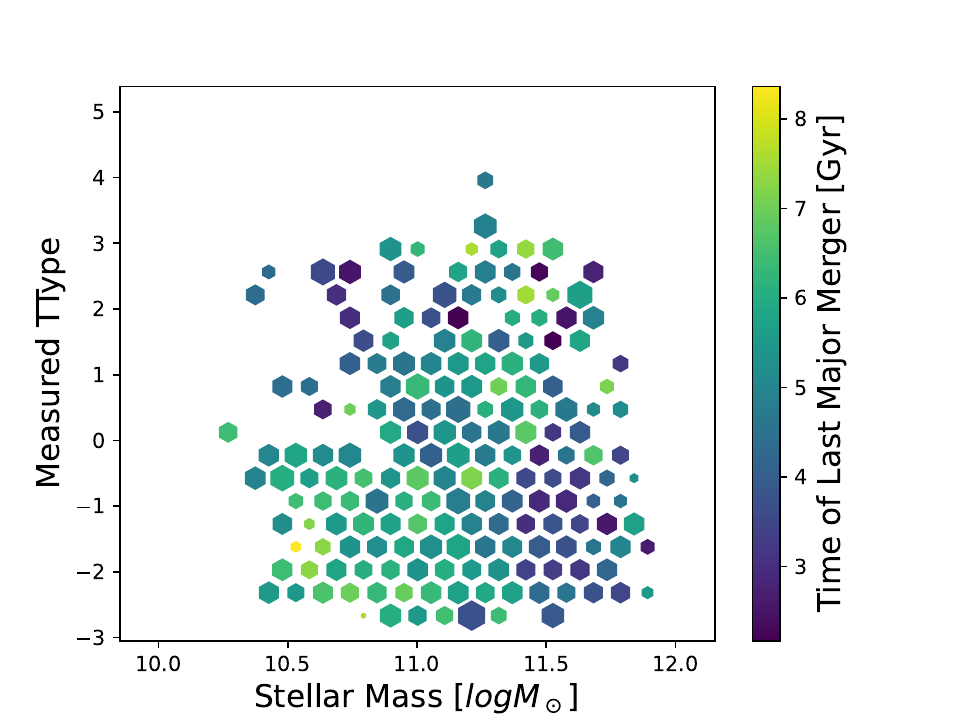}
    \caption{{\bf The ex-situ stellar mass fraction and the mass and time of the last major merger inferred for HSC galaxies of different stellar morphologies.}
     From top to bottom we plot 2D histogram of ex-situ fractions, mass and time of the last major merger, respectively, in the plane of morphological TType classification and galaxy stellar mass. As a reminder, positive TTypes values $>1$ denote disky galaxies, whereas negative ones $<-1$ are for more elliptical systems. Medians, errors and annotations are as in Figures~\ref{fig:inference_exsitu_mass_size} and \ref{fig:inference_mass}. Based on their images, also HSC galaxies seem to support a picture whereby  ellipticals have undergone more massive recent major mergers and have accumulated larger fractions of ex-situ stars than disky ones with similar mass.
    }
    \label{fig:inference_ttype_mass}
\end{figure}

In Figure~\ref{fig:inference_ttype_mass}, we finally show the inferred ex-situ fractions and masses/lookback times of the last major merger plotted against TTypes (morphological classifications) and stellar mass. A strong trend is visible: galaxies with lower TTypes (i.e. elliptical galaxies, around TType $-2$) exhibit significantly higher ex-situ fractions compared to disk galaxies (positive TTypes), which typically have an average ex-situ fraction around 10 per cent. The ex-situ fractions in the TType–stellar mass plane further reveal that, at a given stellar mass, elliptical galaxies tend to have higher ex-situ fractions than disk galaxies. We find a similar trend with TTypes for the mass of the last major merger, although here the prediction errors are smaller towards the more massive end. We also see that, at a given stellar mass, disk galaxies (positive TType) tend to have less massive mergers (or even no major merger at all) compared to their same mass elliptical (negative TTypes) counterparts. This underlines the relevance of major mergers on determining the overall morphology of a galaxy. For the lookback time of the last major merger, trends with TTypes are less clear. This is especially contrasting to the trends with galaxy sizes in Figure~\ref{fig:inference_mass}. It may imply that the recentness of a major merger has a larger impact on the size of a galaxy than on its morphology. 

In Appendix~\ref{sec:prior} we quantify in more detail the level to which the ML model learns possible underlying cross-correlations between properties of the merger and assembly histories of galaxies. In this respect, we also quantify possible differences between the HSC and TNG predictions on a population-wise basis.

\section{Discussion}
\label{sec:discussion}
In this work, we have performed simulation-based inference of aspects of the merger and stellar assembly histories of HSC galaxy images using the outcome of the TNG50 and TNG100 cosmological galaxy simulations as the training set. Based on the validation via the ground truth available for the simulated TNG galaxies, we are able to judge the performance of our model. The inference delivers a high amount of information content, especially for the stellar ex-situ fraction. Additionally, the mass of the last major merger can be well recovered for higher masses above $9.5 \log \MSUN$. Comparing this to our initial approach where the input observables were scalar galaxy properties rather than images \citep{Eisert_2023}, we find that, although the results are similar, the overall errors seem to be larger. How can this be, given that the images should contain more information than the scalars inferred from them? We offer two explanations in response to this question. 

First, the scalars used by \cite{Eisert_2023} were not all truly observable. In particular, the fraction of disc stars, calculated in TNG galaxies using the intrinsic kinematics of stellar particles, is not a directly measurable quantity. Meanwhile, it was shown by \cite{Eisert_2023} that including the kinematic property of stellar particles greatly benefits inference of galaxy ex-situ fractions. In practice, disk stellar mass can be inferred \emph{a posteriori} from multi-wavelength images using component-wise spectral energy distribution fitting \citep[e.g.][]{2022ApJ...930..153S,2024MNRAS.528.5452B} or Schwarzschild modelling with access to spatially-resolved optical spectroscopy  \citep[e.g.][]{2024MNRAS.534..502S}. But neither of these inference techniques, nor our representations, are comparable to having \emph{a priori} knowledge of the intrinsic, dynamical stellar components. Thus, the new predictions presented in this paper may be limited by the extent to which the kinematics of stellar components are encoded in the image-based representations.   

Second, here we have added observable realism to the images. This benefits the domain compatibility between TNG and HSC galaxies, but also introduces significant amount of information loss that is unavoidable in real-world applications. A qualitative visual comparison of pre-realism\footnote{\url{https://www.tng-project.org/explore/gallery/bottrell23i}} and post-realism\footnote{\url{https://www.tng-project.org/explore/gallery/bottrell23}} images of TNG galaxies reveals loss of stellar halo features such as shells and streams that are signposts of ex-situ assembly \citep[e.g.][]{1972ApJ...178..623T,1988ApJ...331..682H,2008ApJ...689..936J}. Characterization of these features is sensitive to the resolution and depth of images \citep[e.g.][]{2023MNRAS.521.3861D,2024MNRAS.528.5558W,2024MNRAS.532..883S,2024MNRAS.tmp.2187B}. This sensitivity also extends to the inner stellar components, such as bulges and disks \citep[e.g.][]{2019MNRAS.486..390B}.  

Despite these limitations, we still conclude that the inference of merger/assembly history statistics based on images is feasible and in fact robust, at least for certain properties. Regarding the domain shift between the dataset we train on (TNG50 and TNG100) and the dataset to which we apply the model, we can report that simultaneous training of the cINN with a log-likelihood loss and the ResNet via contrastive learning/NNCLR keeps the domains significantly overlapping. In the following sections, we will discuss more closely the options to improve the capability of the model and how trustworthy our inference results are.

\subsection{What can be done to further improve the inference results?}
In the preparation of the datasets, we have access to  observable quantities such as redshift, apparent i-band brightness, and Petrosian apparent radius (e.g. to match samples from different domains). During preprocessing of the images, the pixel values were normalized to a range of $[0,1]$, effectively scaling away any information content from the relative brightnesses of galaxies. Similarly, relative size information is suppressed by our scaling of $r_{\rm p90}$ to set the field of view. These normalizations could reduce the model's inference accuracy by removing critical contextual information. For this reason, we decided to re-integrated the redshift, apparent radius, and apparent brightness into the inference model as additional inputs -- the entirety of the results presented in this paper is based on including such additional available observables. Providing these contextualizing characteristics alongside the scaled images leads to significant improvements in the model's performance. Specifically, the uncertainties in e.g. the ex-situ fraction predictions decrease from $\approx \pm 20$ per-cent to $\approx \pm 10$ per-cent. The \emph{relative} context of the images therefore contains crucial information for constraining the merger and assembly histories of galaxies. We theorize that e.g. including statistics describing the environment of a galaxy (which is not visible on the zoomed-in images) might lead to further significant improvements.

We propose the inclusion of additional informative merger/assembly and integral galaxy properties to be inferred from the images alongside the unobservable properties analyzed in this work (i.e., the ex-situ fraction, as well as the mass and lookback time of the last major merger). For instance, significant cross-correlations with ex-situ fractions have been identified in simulation-based studies \citep{Zhu_2022, Davison_2020} for fundamental galaxy properties like stellar mass, and stellar mass was also utilized in our previous scalar-based approach \citep{Eisert_2023}. It would be worthwhile to check whether it is possible to actually infer the very stellar mass of a galaxy from its images rather than requiring additional observable data during the inference phase. 

We also want to comment on the lookback time of the last major merger, which is not well recovered by our current model. This could be interpreted to mean that the current images encode the information that \emph{something} happened to a galaxy, but not \emph{when} it happened. This information could be introduced by observable features that dissipate on characteristic time scales -- such as streams and shells in stellar halos. While we see such features in some HSC images, particularly for bright low-$z$ galaxies, we do not see them ubiquitously. Future surveys, such as the Legacy Survey of Space and Time (LSST) at 10-year depth sensitivity, stand to offer more complete access to these faint, but useful features \citep[e.g. Section 9.6.3 in ][]{LSST_Science_Book}.

We also suggest that incorporating additional spectrophotometric data containing information about stellar populations \citep{Thorne_2021, Sanders_2021} and stellar kinematics \citep{Wisnioski_2015, Tiley_2019} could enhance the predictive capabilities of our inference model. In this work, we already utilize the r-, g-, and i-bands, which could easily be extended by including the z- and y-bands. However, since these bands are shallower in depth compared to the g- and r-bands, it remains to be seen whether their inclusion would add significant value.
In this respect, a potential concern with our current approach is that all channels are stretched and normalized separately, meaning that the relative brightness between the bands is not fully preserved. A workaround for future work could involve generating a normalized combined brightness channel, L, as a composite of all bands, and then creating a residual image for each band by subtracting it from the combined L channel. This approach would enhance the model's ability to focus on the differences between the channels (i.e. colours).

Our model is using multi-broadband optical light. Optical imaging data for galaxies is abundant and we showed that this data alone suffices to infer important characteristics of galaxy assembly histories. However, together with images, resolved stellar kinematic information from integral field units (IFUs) should offer further constraints on assembly \citep[e.g.][]{Angeloudi_2024}. The trade-off is that this significantly reduces the number of galaxies to which the model can be applied. On the other hand, one could incorporate physical properties derived from integrated galaxy spectra \citep[e.g. from BOSS][]{BOSS_1, BOSS_2}, whereas we only provided the spectroscopic redshifts in the model developed here.   

Further useful information about the assembly histories may be available in the distribution of cool gas traced by the 21cm HI fine-structure line. The HI distributions of galaxies are diverse, even among galaxies in low-density environments where stripping is not expected \citep{2023PASA...40...32R}. The distribution of gas in galaxies is affected by mergers \citep{1996ApJ...464..641M}, but also by gas accretion from satellites and diffuse inflows (e.g. \cite{2020MNRAS.491.3908R}). Considering the diversity in how HI is distributed and these links to assembly, it could be expected that HI would improve inference. However, these links are complicated by the roles of supernova, stellar winds, and active galactic nuclei in regulating the distribution of gas in-and-around galaxies \citep{2023MNRAS.524.5391A}. Nonetheless, incorporating information about the cool gas is appealing, considering that it can be measured for large galaxy samples by HI surveys such as WALLABY using the Australian Square Kilometre Array Precursor facility \citep{2020Ap&SS.365..118K}.

Finally, it is important to consider how the uncertainty in Figure~\ref{fig:results/TNG100/posterior_example} is formed: by the intrinsic information limitation of the input, insufficient realism of the simulation, or insufficient modeling? We look into this in the next section.

\subsection{Can we trust the inference?}
By using contrastive learning that includes both simulated and observed galaxy images, we aim to ensure that the model does not overly focus on simulation-specific features. This approach is proven largely successful, as we were able to transfer information from the TNG simulations onto the HSC data. However, we recognize that our mock images are not perfect, and there are both small and large deviations between the simulated and observed domains. As in previous work, the definition of similarity remains somewhat arbitrary, and it is not guaranteed that two visually identical galaxy images represent identical physical objects.

A critical aspect to consider is the quality of the matching between the simulations and observations. Our current approach proved to be useful and appropriate to identify non-similar galaxies \citep{Eisert_2024}. This said, the nearest neighbor search that our OOD score is currently based on is in high-dimensional spaces only sensitive to very strong changes in actual point densities. This is worsened by the sparsity in data points in high dimensions. To improve our understanding of the alignment between datasets, we could therefore also employ a normalizing flow to learn the distribution in the representation space. A continuous representation learned in this manner would allow us to calculate the probability of finding a galaxy of a certain type within the datasets. Alternatively, or additionally, we could also run the score calculation on a lower-dimensional representation, such as a UMAP, to further explore the alignment and relationships between the datasets. 

All this considered and and given the OODs, how trustworthy are the results if the datasets are not 100 per cent aligned in representation space? 

To this aim, we have extended our analysis to include the EAGLE simulation, by generating HSC-like EAGLE mocks using the same pipeline that was used to create the TNG mocks.
The results in Figure~\ref{fig:map_eagle} showed that the model performs well on EAGLE data, with no significant deviations in inference results compared to those obtained from TNG. This success is notable because EAGLE was matched against the HSC population, even though our model is trained on TNG. Additionally, the OOD scores for the EAGLE/TNG comparison can extend above the values 1.2, indicating significant differences between the images from EAGLE and TNG. This suggests that our model is capable of effectively extrapolating beyond the TNG dataset. However, this capability might become reduced when applied to higher-redshift data, such as those from LSST or HSC-DEEP, where finer image details become more critical.

The success of our TNG-trained inference model on EAGLE demonstrates that the model is capable of accurate inference across different simulation datasets, reducing concerns about potential over-fitting to TNG-specific features. However, a crucial step for further validation would be a comparison with other inference methods, such as the MANGA results of \cite{Angeloudi_2024}. Given that we use different samples, it would be necessary to ensure that we are comparing inference results for the same galaxies. This should likely restrict the application of our SBI model to SDSS galaxies with sufficient MANGA observations, but there should be a significant overlap between the MANGA dataset and the HSC set we are using. This seems a promising and important avenue for future work.

\subsection{Stellar ex-situ fractions of HSC galaxies}
Our SBI utilizes a sample of simulated TNG50 and TNG100 galaxies that have been matched to observed HSC galaxies using certain observables (Table~\ref{tab:observable_properties}). Furthermore, we removed galaxy images in both sets with OOD scores $> 1.2$. This selection process inherently filters out certain types of galaxies, including potentially blue disks, those with larger physical radii, higher asymmetries, and greater concentrations. More details on the relation between OOD value and observable properties have been discussed by \cite{Eisert_2024}.

Despite these constraints, our model successfully reconstructs fundamental relationships found in simulations, such as the well-established ex-situ stellar mass relation \citep{Pillepich_2018} and the ex-situ vs. galaxy size relation \citep{Zhu_2022, Davison_2020}, according to which more massive or larger galaxies build up their stellar mass primarily via mergers and stellar mass accretion. Additionally, the morphological galaxy type seems very strongly connected to the ex-situ fraction \citep{rodriguezgomez2016role}.

Even more fundamentally, we find that according to HSC images, i.e. real galaxies, the ex-situ fraction increases with stellar mass, particularly for galaxies with masses above $10^{10.5} , \MSUN$. This trend is also observed in more extended galaxies, with low TTypes exhibiting particularly high fractions of accreted stars (lenticular/elliptical morphologies). These findings align with the theoretical understanding that more massive and spheroidal galaxies tend to accumulate a larger portion of their stellar mass from external sources through mergers \citep{Tacchella_2019}.

However, an intriguing outcome of our SBI application is that the median ex-situ fraction versus mass relation inferred for HSC, which is trained on the TNG50/TNG100 simulations, differs somewhat from the theoretical TNG relation. This discrepancy raises important questions about the factors influencing our model's predictions. By training our model on TNG, we inherently take the TNG relation as a prior. In scenarios where the data are sparse or of lower quality, we might expect the model to simply sample this prior. For instance, if we provided the model with images containing only sky noise, it would likely fail to recover any meaningful trends with mass, leading to an ex-situ fraction distribution that is independent of mass.

In contrast, our model recovers trends with mass, indicating that the galaxy data is sufficiently informative to overcome the prior assumptions embedded in the TNG relation.

The lower median ex-situ fractions in HSC compared to TNG may imply that there are more galaxies at intermediate stellar masses with lower ex-situ fractions in HSC, and possibly in general inthe Universe, than in TNG. These galaxies may be bluer and more disk-like than their TNG counterparts, indicating that the HSC dataset encompasses a different population of galaxies that may not be fully captured by the TNG simulations.

In future studies, we plan to investigate these differences in greater detail. A first step forward would be to extend our model to also infer galaxy properties such as stellar mass, diskiness, and color along the merger/assembly history. This will enable us to analyze these relationships independently from the properties traditionally determined via observational methods, as done in the previous HSC figures. In other words, for this study, beyond the three merger and stellar assembly properties inferred in this work, we have limited our analysis to catalog data, as inferring additional galaxy properties would require extensive validation. It might be that the difference in the median ex-situ fractions at fixed galaxy stellar mass inferred for HCS in comparison to TNG data may actually be due to differences in the underlying values of galaxy stellar masses on the x-axes.

At this point, we can compare our results to the study done by \cite{Angeloudi_2024}. While their ex-situ inference model is based on MANGA images from the local Universe, both studies use results from TNG as training data and rely on the same definition of ex-situ stars. In that work, the trend in higher ex-situ fractions at fixed mass for different morphologies is also seen. Interestingly, we notice that in their case the inferred ex-situ fraction versus mass relation is above the TNG50 ground truth, rather than below as found here. A reason may be that our simulations sample is dominated by TNG100 galaxies that are in average more massive. Furthermore this suggests that our sample might also contain a higher proportion of elliptical galaxies with higher ex-situ fraction compared to the MANGA sample, potentially explaining the observed difference.

\section{Summary and Conclusions}
In this paper, we have performed simulation-based inference of the assembly and merger histories of galaxies observed with the Hyper Suprime-Cam (HSC) survey using optical HSC g, r, and i-band galaxy images as input. Our model was trained on mock images from the TNG50 and TNG100 cosmological simulations and was also provided with spectroscopic redshifts, $i$-band apparent brightnesses, and angular sizes as inputs alongside the images.

We focused on inferring posterior distributions for key properties such as the ex-situ stellar mass fraction, and the mass and lookback time of the last major merger for each galaxy. Our methodology combines a ResNet architecture trained via self-supervised contrastive learning on both TNG and HSC galaxies, and a conditional Invertible Neural Network (cINN) trained on the ground truth data from the TNG simulations. The galaxy samples span a stellar mass range of $10^{9-12}\, \mathrm{M}_\odot$ and redshifts between $z=0.1$ and $z=0.4$.

\subsection*{Inference validation}
Validation against the TNG ground truth demonstrates the model's robust performance (Figure~\ref{fig:results/TNG100/map}):

\begin{itemize}
    \item The ex-situ stellar mass fractions achieve prediction errors within $\pm 10$ per-cent for $80$ per-cent of the sample, enabling clear differentiation between low and high ex-situ fraction galaxies.
    \item The mass of the last major merger is accurately recovered for masses above $10^{9.5}\, \mathrm{M}_\odot$, with prediction errors around $\pm 0.5$ dex for $80$ per-cent of the sample.
    \item The lookback time of the last major merger, on the other hand, remains more challenging to be constrained precisely.
\end{itemize}

Comparing our image-based approach to previous work utilizing scalar properties \citep{Eisert_2023}, we achieve less accurate results, chiefly because of the increased observational realism of the images compared to the idealized scalars. Additionally the images alone undergo information loss due to the normalization in brightness and field of view: we have counteracted this by adding the redshift, apparent galaxy sizes and apparent luminosities as additional scalar inputs alongside the images.   

To assess the generalizability of our procedure across different cosmological simulations, we have applied it to mock images from the EAGLE and SIMBA simulations, processed through the same image creation pipeline as TNG:

\begin{itemize}
    \item We compare the alignment of the survey-realistic images generated from TNG/EAGLE/SIMBA in the representation space obtained via contrastive learning. By this we find that TNG and EAGLE galaxies are more similar to each other than to SIMBA galaxies. Furthermore we find that the overall galaxy populations of TNG, followed by EAGLE's, are much more similar to HSC than to SIMBA (Figures~\ref{fig:umaps}, \ref{fig:similarity_distribution_pre_matched}, \ref{fig:similarity_distribution}, and \ref{fig:umaps_2}).
    \item Training on TNG  and testing on EAGLE returns similarly goodinference accuracy as training and testing on TNG images only, underscoring the robustness of our simulation-based inference and ability to transfer learned knowledge across different simulated universes with distinct subgrid physics (Figure~\ref{fig:map_eagle}).
\end{itemize}

\subsection*{The merger and assembly histories of HSC galaxies}
We applied our model to observed HSC images, providing the community with inferred ex-situ stellar mass fractions and masses and times of the last major merger for more than 750'000 real galaxies, which we hereby make publicly available. 

We discover the following connections in real HSC data:
\begin{itemize}
    \item Ex-situ stellar mass fractions increase with both galaxy stellar mass and size, with the average HSC galaxies showing systematically lower ex-situ fractions by about 10–15 per-cent than simulated TNG galaxies at fixed stellar mass, especially between $10^{10.5}$ and $10^{11.5} \MSUN$. Larger galaxies also exhibit higher ex-situ fractions even at fixed mass (Section \ref{sec:result_1}). 
    
    \item More massive galaxies experienced more massive last major mergers, but the merger timing shows no clear dependence on stellar mass, with HSC galaxies having slightly earlier mergers (by 1–2 Gyr) than TNG galaxies at $\approx 10^{10.5-11.2} \MSUN$. Among HSC galaxies, ellipticals, in contrast to disks, show strong secondary trends: at fixed stellar mass, larger ellipticals have undergone more massive and more recent major mergers, consistent with their higher ex-situ stellar mass fractions (Section \ref{sec:result_2}).
    
    \item Galaxies with lower TTypes (ellipticals) have significantly higher ex-situ stellar mass fractions and more massive last major mergers than disk galaxies of similar stellar mass, highlighting the key role of mergers in shaping the morphology of galaxies. In contrast, the timing of the last major merger shows a weaker correlation with TType, suggesting that merger mass influences morphology more strongly than the time of a merger (Section \ref{sec:result_3}).
\end{itemize}

Our findings demonstrate that simulation-based inference using galaxy images is not only feasible, but also highly informative, enabling the extraction of previously unobservable merger and assembly history details from observational data. This approach offers a powerful tool for bridging simulations and observations, in principle allowing for more nuanced comparisons and enhanced understanding of galaxy evolution processes.

\section*{Acknowledgements}
LE and this work are supported by the Deutsche Forschungsgemeinschaft (DFG, German Research Foundation) under Germany’s Excellence Strategy EXC 2181/1-390900948, Exploratory project EP 3.4 (the Heidelberg STRUCTURES Excellence Cluster). CB gratefully acknowledges support from the Forrest Research Foundation. DN acknowledges funding from the Deutsche Forschungsgemeinschaft (DFG) through an Emmy Noether Research Group (grant number NE 2441/1-1). RSK acknowledges financial support from the ERC via Synergy Grant ``ECOGAL'' (project ID 855130),  from the German Excellence Strategy via the Heidelberg Cluster ``STRUCTURES'' (EXC 2181 - 390900948), from the German Ministry for Economic Affairs and Climate Action in project ``MAINN'' (funding ID 50OO2206), and from DFG and ANR for project ``STARCLUSTERS'' (funding ID KL 1358/22-1). The primary TNG simulations were carried out with compute time granted by the Gauss Centre for Supercomputing (GCS) under Large-Scale Projects GCS-ILLU and GCS-DWAR on the GCS share of the supercomputer Hazel Hen at the High Performance Computing Center Stuttgart (HLRS). Part of this research was carried out using the High Performance Computing resources at the Max Planck Computing and Data Facility (MPCDF) in Garching, operated by the Max Planck Society. We thank the Max Planck computing support teams for their valued assistance. 

The Hyper Suprime-Cam (HSC) collaboration includes the astronomical communities of Japan and Taiwan, and Princeton University. The HSC instrumentation and software were developed by the National Astronomical Observatory of Japan (NAOJ), the Kavli Institute for the Physics and Mathematics of the Universe (Kavli IPMU), the University of Tokyo, the High Energy Accelerator Research Organization (KEK), the Academia Sinica Institute for Astronomy and Astrophysics in Taiwan (ASIAA), and Princeton University. Funding was contributed by the FIRST program from the Japanese Cabinet Office, the Ministry of Education, Culture, Sports, Science and Technology (MEXT), the Japan Society for the Promotion of Science (JSPS), Japan Science and Technology Agency (JST), the Toray Science Foundation, NAOJ, Kavli IPMU, KEK, ASIAA, and Princeton University. This paper makes use of software developed for Vera C. Rubin Observatory. We thank the Rubin Observatory for making their code available as free software at \url{http://pipelines.lsst.io/}. This paper is based on data collected at the Subaru Telescope and retrieved from the HSC data archive system, which is operated by the Subaru Telescope and Astronomy Data Center (ADC) at NAOJ. Data analysis was in part carried out with the cooperation of Center for Computational Astrophysics (CfCA), NAOJ. We are honoured and grateful for the opportunity of observing the Universe from Maunakea, which has the cultural, historical and natural significance in Hawaii. 

\section*{Data Availability}

Data directly related to this publication and its figures will be made available on request from the corresponding author. In fact, the IllustrisTNG simulations are already publicly available and accessible in their entirety at \href{www.tng-project.org/data}{www.tng-project.org/data} \citep{nelson2019illustristng}. The analysis code and ML models developed here are also available upon request. The data pertaining the survey-realistic mocks of TNG50 and TNG100 galaxies have been made available on the same IllustrisTNG website \citep{Bottrell_2023}. With this paper, we also release the survey-realistic mocks of EAGLE and SIMBA galaxies at $z=0$ and a catalog with the merger and assembly properties inferred for more than 700`000 HSC galaxies.



\bibliographystyle{mnras}
\bibliography{references} 




\appendix

\section{Calibration Errors}
\label{sec:calibrationerrors}
While the main text of this paper primarily focuses on evaluating the performance of the cINN in terms of Maximum A Posteriori (MAP) estimates and posterior standard deviations, here we use a more quantitative analysis to assess how well the inferred posterior distribution aligns with the ground truth distribution of the galaxy sample. To achieve this, we calculate the calibration error for each target quantity, which reflects the difference between the ground truth distribution from the combined TNG100 and TNG50 galaxy test sample and the corresponding inferred cINN posterior distribution.

The calibration error is determined by evaluating, across the entire sample, the frequency with which the ground truth values fall within specific confidence intervals of the cINN posteriors for each target quantity. For example, if the $50$ per cent confidence interval of the sampled posterior $p(x|c)$ is accurate, we would expect $50$ per cent of the ground truth test galaxies to fall within this interval. A positive calibration error indicates an under-confident posterior (where the interval contains more test galaxies than suggested by the posterior), while a negative calibration error signifies an over-confident posterior (where the interval contains fewer test galaxies than suggested).

\begin{figure}
	\centering
	\includegraphics[width=9cm]{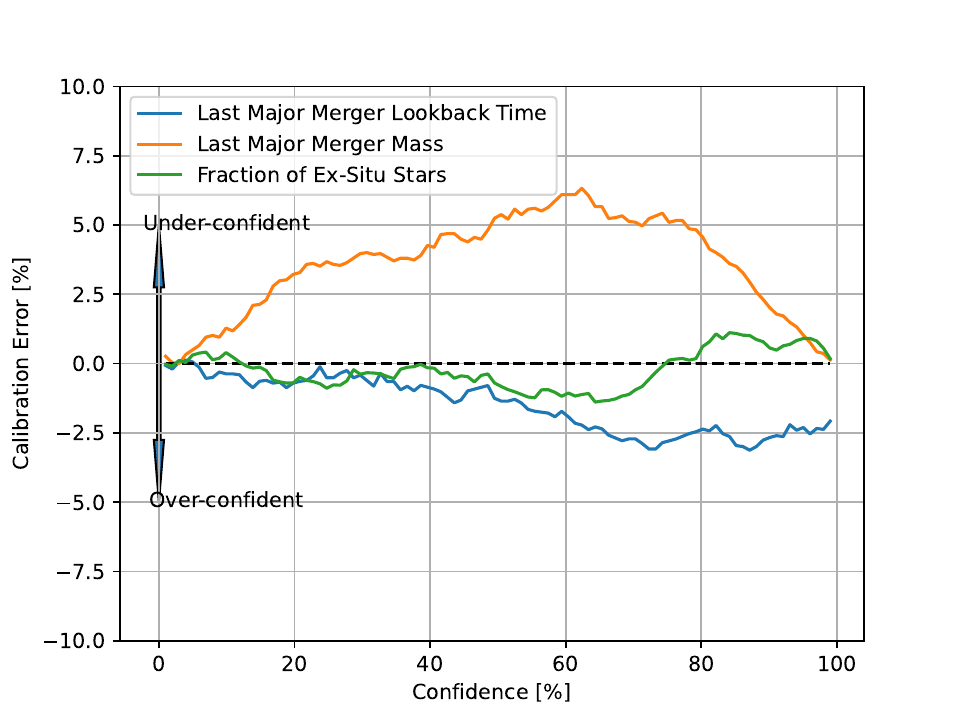}
	\caption{{\bf Calibration error of the cINN model as a function of the confidence level for each of the three target quantities.} The calibration error represents the deviation between the fraction of test galaxy ground truths within a given posterior confidence interval and the expected fraction based on the respective confidence level. A positive calibration error indicates an under-confident posterior, while a negative value denotes an over-confident posterior. We calculate the calibration error based on the TNG50 and TNG100 galaxies in the test set. The ex-situ fraction posteriors exhibit minimal calibration errors, deviating only slightly from the ideal line. The mass of the last major merger shows a tendency towards under-confidence.}
	\label{fig:results/calibration_error}
\end{figure}

Figure~\ref{fig:results/calibration_error} illustrates the calibration errors as a function of the posterior confidence level for each of the three outputs. These calculations are conducted on the raw posteriors without applying any additional binning, such as the previously-used "No-Major-Merger" bin. The ex-situ fraction displays a well-behaved posterior with calibration errors remaining within a few percent of the ideal line. The mass of the last major merger is found to be slightly under-confident.

\section{Similarity of image samples across training efforts}
\label{sec:OODsafterretraining}

\begin{figure*}
	\centering
 \includegraphics[trim={0 0.3cm 0 1cm},clip,width=\linewidth]{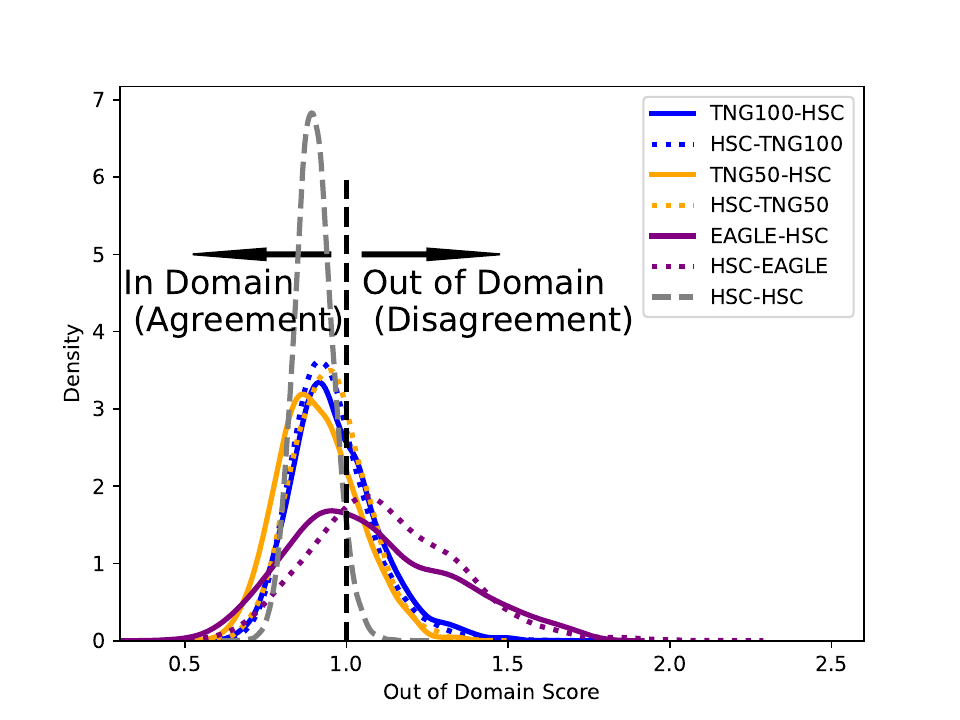}
	\caption{{\bf Distributions of the Out of Domain Score (OOD) after training the pre-trained ResNet for the contrastive learning step together with a cINN.} The plot shows the OOD score distributions after excluding galaxies with OOD > 1.2 and retraining the ResNet from \citet{Eisert_2024} with a cINN head to infer merger and assembly history properties (see Section~\ref{sec:pipeline}). This analysis demonstrates that the retraining of the ResNet during the inference does not cause a significant domain shift between simulations and observations. The results should be compared to Figure 10 in \citet{Eisert_2024} for TNG and to Figure~\ref{fig:similarity_distribution_pre_matched}.}
	\label{fig:similarity_distribution}
\end{figure*}

In Figure~\ref{fig:similarity_distribution} we provide the comparison of the OOD score distributions of all the simulated galaxy samples and the observed ones that are obtained by retraining the contrastive learning ResNet in tandem with the cINN for the inference. These distributions shall be compared to those of Figure~\ref{fig:similarity_distribution_pre_matched}. 

For the EAGLE simulated galaxies, we find that, even after retraining the ResNet model, the OOD scores do not significantly worsen, suggesting that EAGLE galaxies remain within a reasonable proximity to the observed HSC data. This finding allows us to proceed with testing our TNG-trained model on EAGLE galaxies.

We also show the analog of Figure~\ref{fig:umaps} for TNG100 and TNG50 galaxies by using the representations of their images from the simultaneous training of the Resnet and the cINN: the representations remain well aligned.

\begin{figure*}
	\centering
	\includegraphics[trim={3cm 2cm 1.5cm 1.5cm},clip,width=0.49\linewidth]{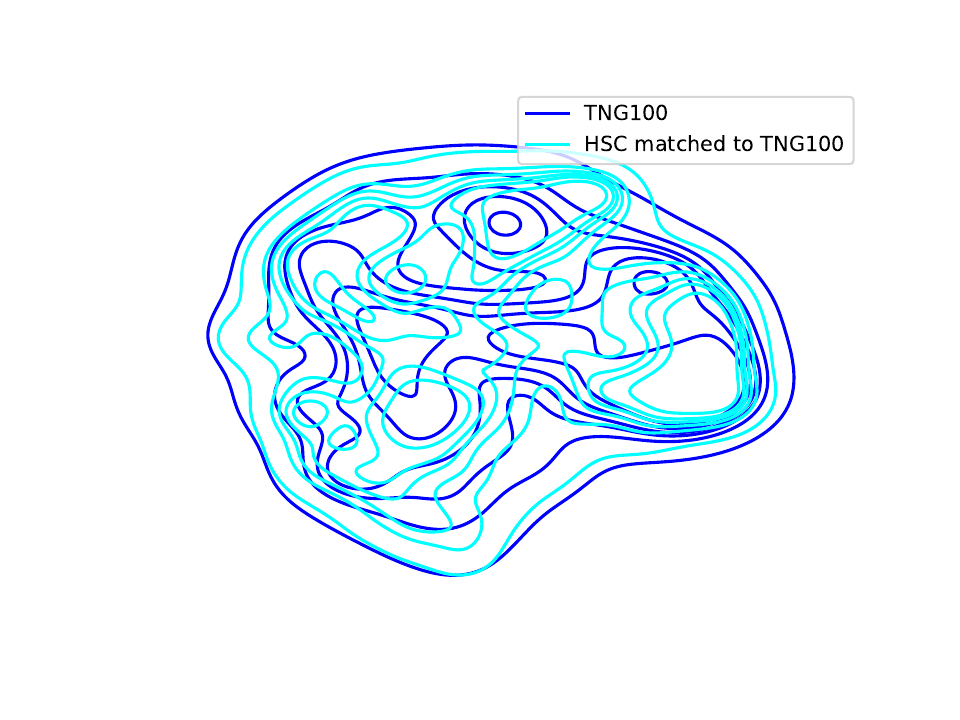}
    \includegraphics[trim={3cm 2cm 1.5cm 1.5cm},clip,width=0.49\linewidth]{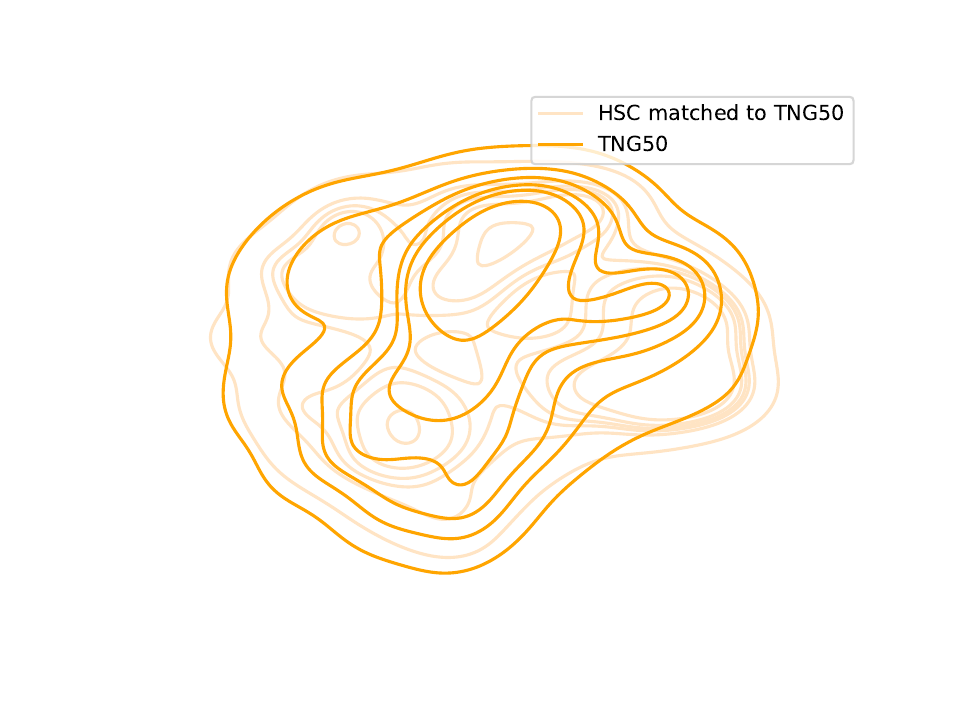}
\caption{{\bf How well do the representations of the observed and simulated galaxy images align to each other, after training the pre-trained ResNet for the contrastive learning step together with a cINN?} We show the 2D UMAP projections of the TNG50 and TNG100 image representations and their respective matched HSC samples like in Figure~\ref{fig:umaps}. However, here we show the distributions after removing galaxy images with a large OOD > 1.2 and by training the ResNet and the cINN at the same time. Note that we don't use the same UMAP representation as in Figure~\ref{fig:umaps}.}
	\label{fig:umaps_2}
\end{figure*}

\section{Cross-correlations between Merger Statistics} 
\label{sec:prior}

\begin{figure}
	\centering
	\includegraphics[width=11cm]{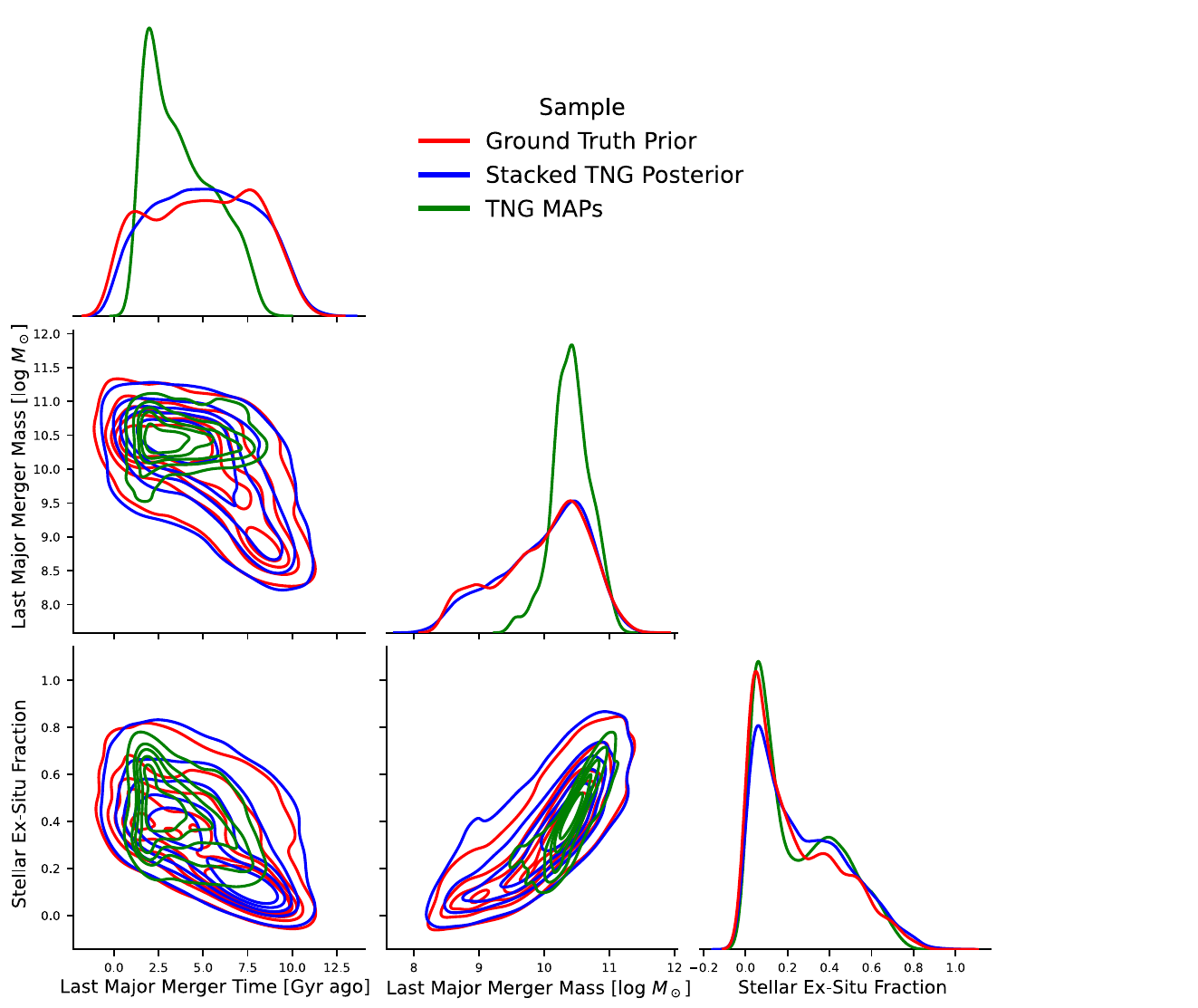}
	\caption{{\bf Has our ML model learned the cross-correlations between the merger statistics?} We show the distribution of the time and mass of the last major merger and the ex-situ fraction on the diagonal. In the lower-left edge, we display the respective cross-correlations. The distributions are plotted for all TNG50 and TNG100 images in our test set. The ground truth from the simulation is shown in red, the stacked posterior samples inferred by our model in blue, and the corresponding MAPs in green. Galaxies without any major merger in their lifetime are excluded. The stacked posteriors follow the ground truth distribution well, while the distribution in MAPs follows the ground truth distribution only for the ex-situ fraction.}
	\label{fig:results/prior}
\end{figure}

We aim to ensure that not only are the predictions for each merger statistic accurate, but also that the cross-correlations between the statistics are preserved. To assess this, we plot in Figure~\ref{fig:results/prior} the ground truth (prior), the stacked posterior (which ideally should resemble the prior), and the MAPs (i.e., the modal value of the posterior).

Overall, we find that the stacked posterior closely follows the prior, including in the cross-correlation plots. However, the posteriors appear to be slightly smoothed out compared to the ground truth. This is particularly noticeable for the ex-situ fraction, where the posteriors are unable to fully recover the sharp peak of galaxies with almost no ex-situ fractions. This smoothing effect could be an artifact caused by the Gaussian noise (variance of $1$ per-cent) introduced during training and by the fact that this plot is based on only approximately 6,000 test galaxies. Additionally, we excluded galaxies with no major mergers in their lifetime, potentially omitting more low ex-situ galaxies from the analysis.

We also find a strong positive correlation between the ex-situ fraction and the mass of the last major merger, as to be expected given the results of the cosmological simulations. However, the MAPs only capture this relation for larger masses above $10^{10} \MSUN$. Below $10^{10} \MSUN$, the relationship flattens, and the MAPs frequently fall into the No-Major-Merger bin. We also see a weaker anti-correlation between the lookback time and the ex-situ fraction. Here, the MAPs tend to cluster around lookback times of approximately 2.5 Gyr ago and do not encode a strong relation with the ex-situ fraction. A similar pattern is visible in the relationship between the lookback time and the mass of the last major merger: The MAPs primarily cover galaxies with a last major merger mass around $10^{10.5} \MSUN$. Despite the poor accuracy in predicting the lookback time of the last major merger, as shown in Figure~\ref{fig:results/TNG100/map}, the model still captures the overall distribution from the galaxies trained on.

\begin{figure}
	\centering
	\includegraphics[width=11cm]{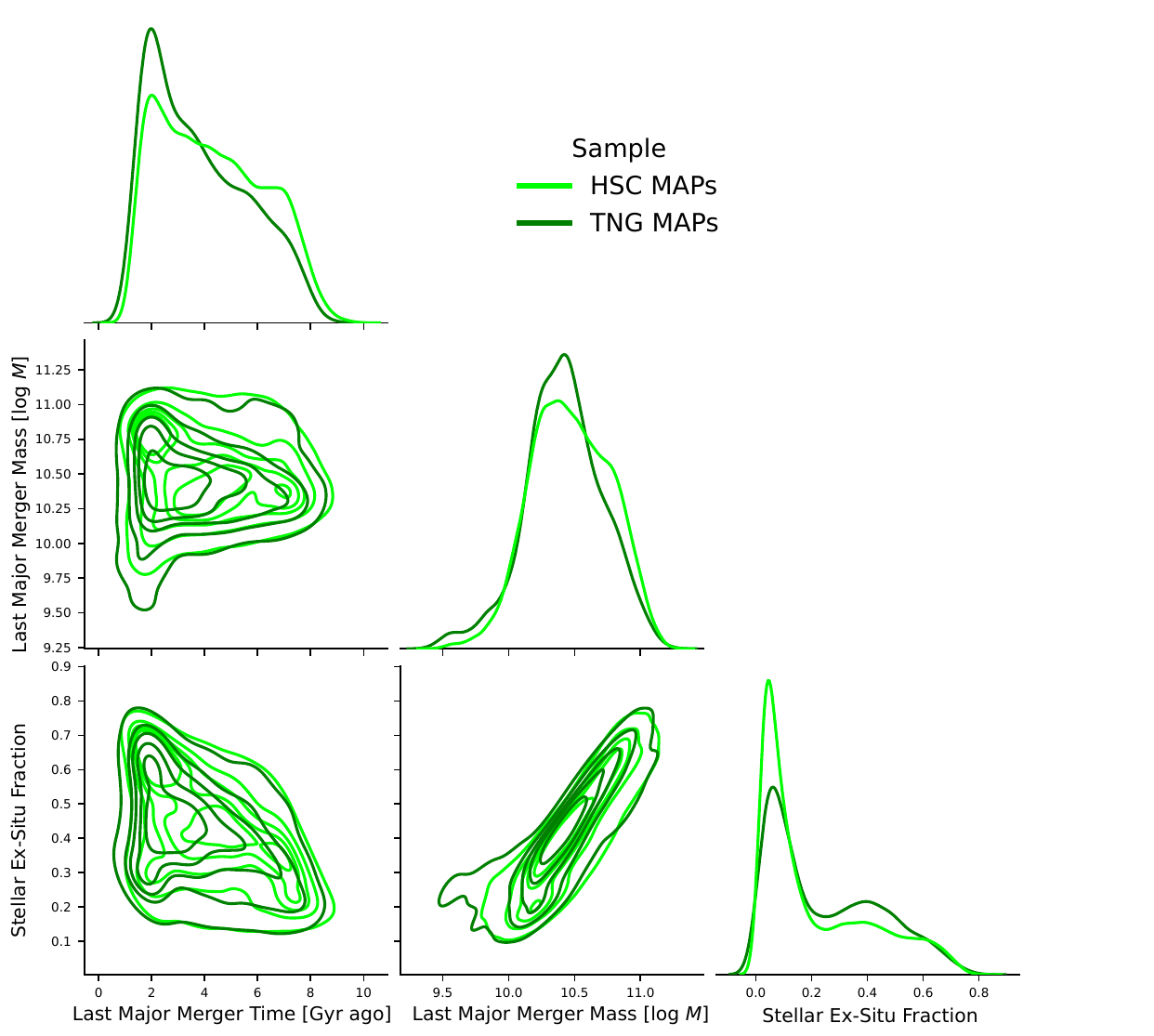}
    \includegraphics[width=11cm]{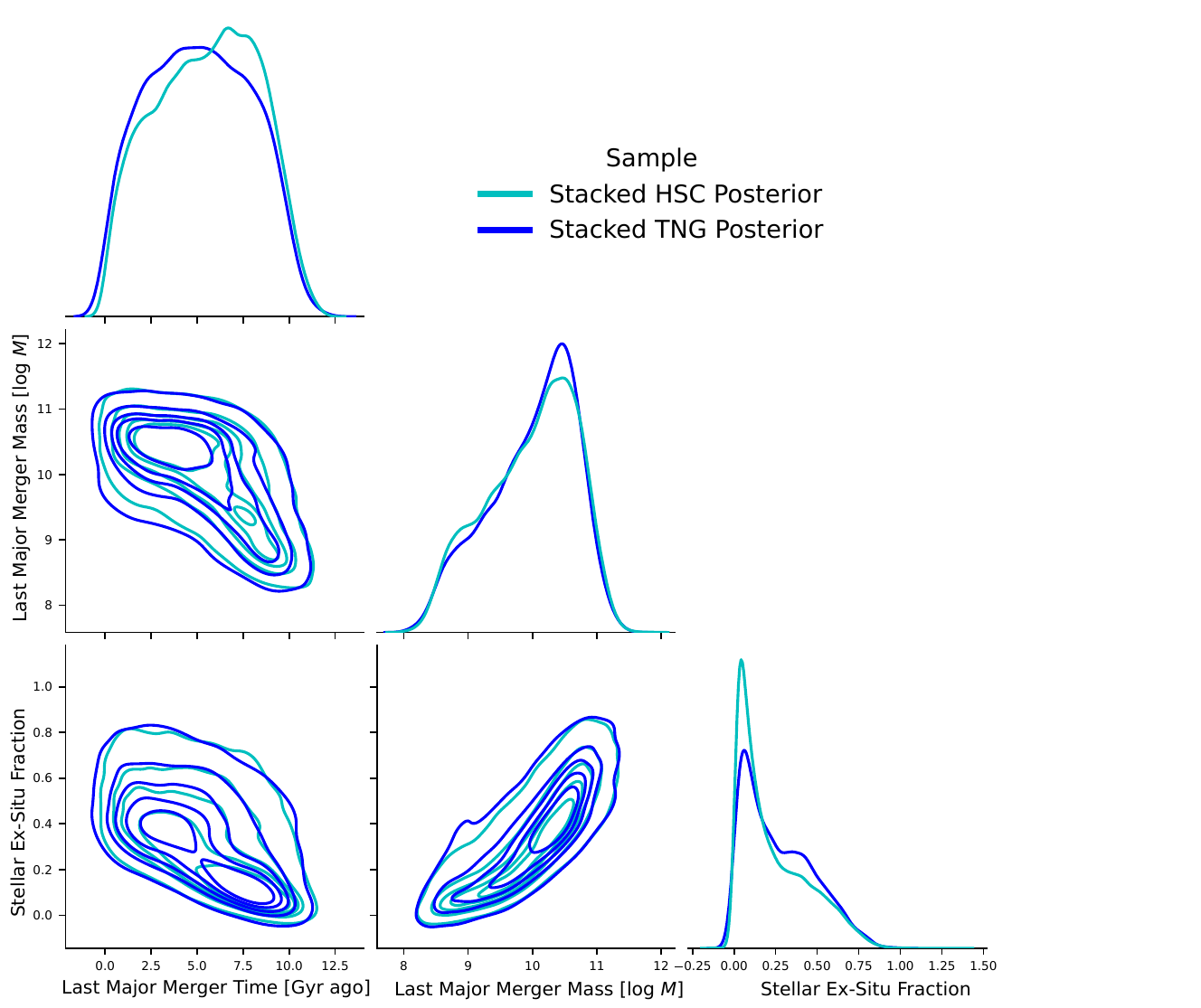}
    \caption{{\bf How do the predictions and their cross-correlations differ population-wise between the simulations (TNG) and the observations (HSC)?} As in the previous figure, we show corner plots of the time and mass of the last major merger and the ex-situ fraction and their cross-correlations. However, here we compare the results between all TNG50 and TNG100 images in our test set with the results for the HSC observed galaxies. In the top panels, we show MAPs (TNG in green and the inferred MAPs for HSC in lime). In the bottom panels, we include all posterior samples (blue for TNG, cyan for HSC). Galaxies without any major merger in their lifetime are excluded.}
	\label{fig:results/prior_2_3}
\end{figure}

We study these cross-correlations by also including now the actual predictions obtained for the observed HSC galaxies and compare them against the predictions of TNG galaxies: see Figure~\ref{fig:results/prior_2_3} for the MAPs (top) and the stacked posterior distributions (bottom). Overall, the relations seem to be well imprinted from TNG to the HSC galaxies. The most intriguing difference is the much larger number of low ex-situ fraction galaxies in HSC. Our model infers that around 60 percent of the HSC galaxies in our sample had no major merger in their lifetime (stellar mass ratio > 1/4 and stellar mass above $8.5 \log \MSUN$). The inference for the HSC galaxies tends slightly toward larger lookback times and higher merger masses. This is an interesting result, considering that mergers with higher lookback times generally have lower merger masses.

\section{Uncertainties of the cINN Predictions} 
\label{sec:uncertainties}
\begin{figure*}
	\centering
	\includegraphics[width=18cm]{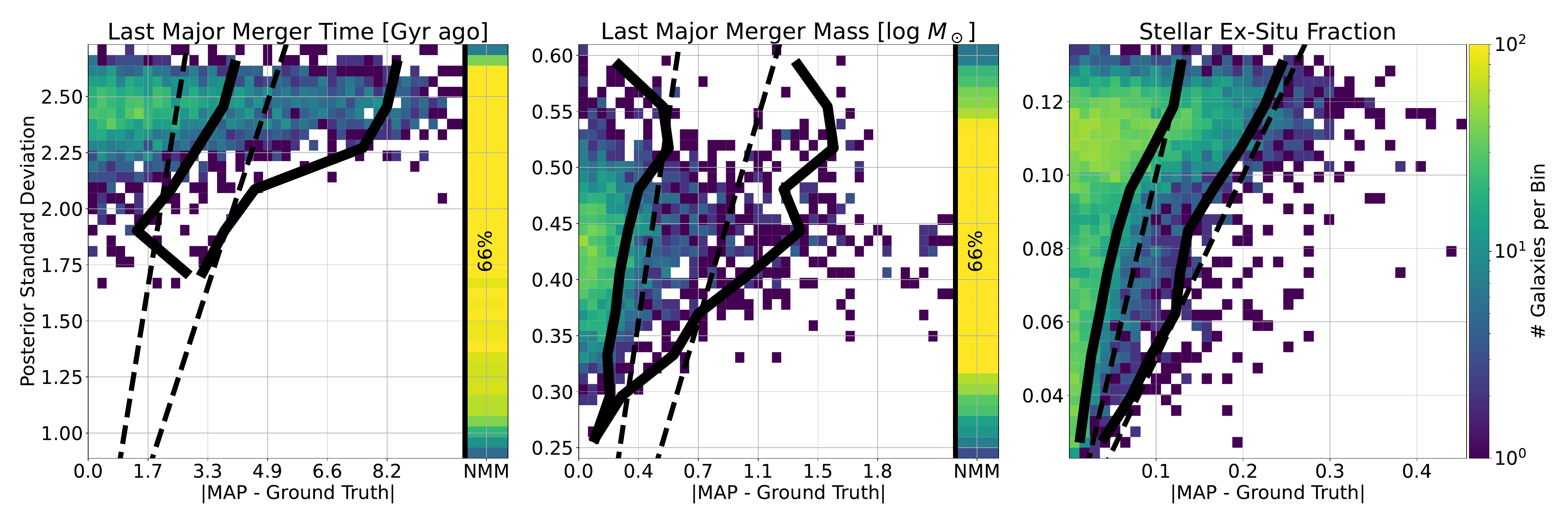}	
    \includegraphics[width=18cm]{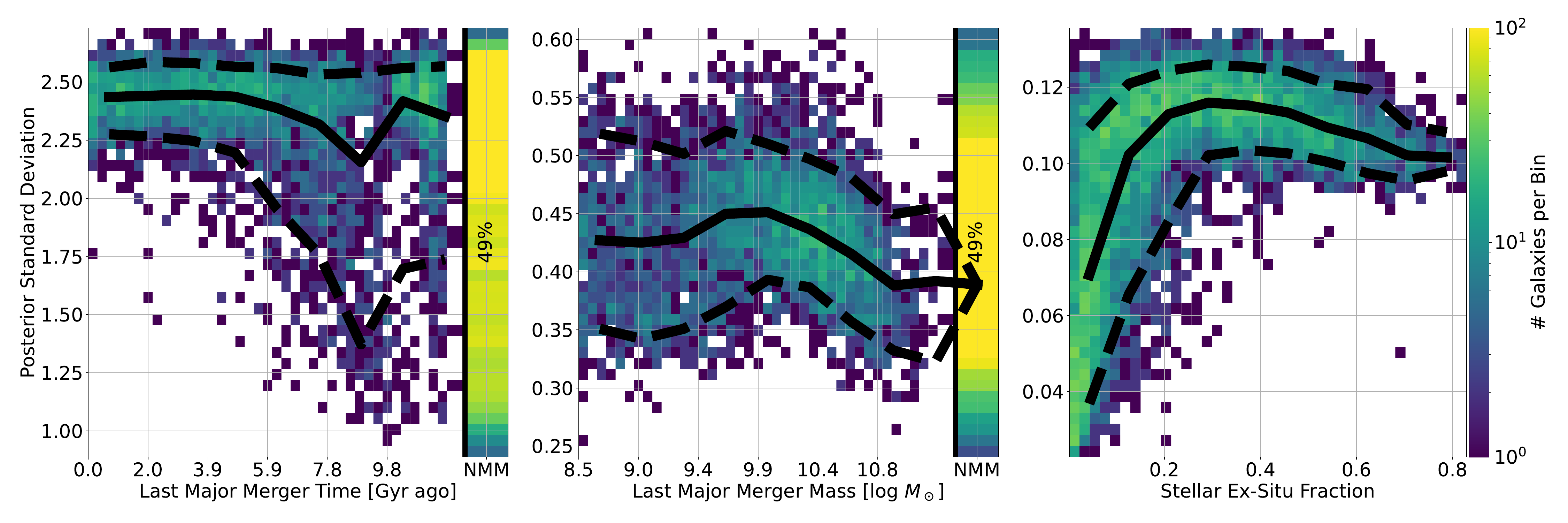}
	\caption{{\bf Uncertainties of the cINN Predictions.}
    {\bf Top:} Two-dimensional histograms of the standard deviation of cINN posterior distributions (y-axis) versus the absolute value of MAP prediction errors (|MAP predictions - ground truth|, x-axis) across all TNG50 and TNG100 test galaxies. Thick black curves represent the 68th and 95th percentiles of test galaxies within bins of posterior standard deviation. Dashed lines illustrate the expected outcome under the assumption of normal distributed posteriors.
    {\bf Bottom:} Two-dimensional histogram of the standard deviation of cINN posterior distributions (y-axis) versus the simulation ground truth (x-axis). The black curves indicate the median, with 80 per cent of test galaxies falling within the area enclosed by dashed lines. For the time and mass of the last major merger, galaxies identified by the ground truth as never having experienced a major merger (NMM) are represented in an additional bin. The fraction of galaxies falling into this bin is denoted. In such cases, standard deviations are calculated solely from posterior samples within the physical range (i.e., lookback time $\le 13.7$ Gyr and stellar mass $\ge 10^{8.5}, \MSUN$).}
	\label{fig:results/TNG100/sigma}
\end{figure*}

To incorporate the information contained in the posteriors, we display in Figure~\ref{fig:results/TNG100/sigma} the standard deviation of the posteriors against the ground truth values and the respective absolute errors of the MAPs (MAP - Ground Truth). Overall, we observe that the qualitative trends are similar to those observed when using a set of scalar observables \citep{Eisert_2023}: the posteriors' widths for the ex-situ fraction are very low for low ex-situ galaxies and increase towards higher ex-situ fractions.




\bsp	
\label{lastpage}
\end{document}